\documentclass[pre,twocolumn,showpacs,preprintnumbers,reprint,amsmath,
                amssymb,superscriptaddress]{revtex4-1}
\usepackage{graphicx} 
\usepackage{dcolumn}  
\usepackage{bm}       
\usepackage{pstricks}

\begin{document}

\newcommand{\bea}{\begin{eqnarray}}
\newcommand{\eea}{  \end{eqnarray}}
\newcommand{\bit}{\begin{itemize}}
\newcommand{\eit}{  \end{itemize}}

\newcommand{\be}{\begin{equation}}
\newcommand{\ee}{\end{equation}}
\newcommand{\ra}{\rangle}
\newcommand{\la}{\langle}
\newcommand{\U}{\widetilde{U}}

\newrgbcolor{fbkblue}{0.2304 0.3476  0.5937}
\newrgbcolor{pksblue}{0.137 0.298 0.513}
\newrgbcolor{verde}{0.267 0.637 0.492}  
\newrgbcolor{verde2}{0.168 0.582 0.543} 
\newcommand{\fbk}[1]{{\color{fbkblue} #1}}


\def\bra#1{{\langle#1|}}
\def\ket#1{{|#1\rangle}}
\def\bracket#1#2{{\langle#1|#2\rangle}}
\def\inner#1#2{{\langle#1|#2\rangle}}
\def\expect#1{{\langle#1\rangle}}
\def\e{{\rm e}}
\def\proj{{\hat{\cal P}}}
\def\tr{{\rm Tr}}
\def\H{{\hat H}}
\def\Hdag{{\hat H}^\dagger}
\def\Lop{{\cal L}}
\def\Ehat{{\hat E}}
\def\Edag{{\hat E}^\dagger}
\def\Shat{\hat{S}}
\def\Sdag{{\hat S}^\dagger}
\def\Ahat{{\hat A}}
\def\Adag{{\hat A}^\dagger}
\def\U{{\hat U}}
\def\Udag{{\hat U}^\dagger}
\def\Zhat{{\hat Z}}
\def\Phat{{\hat P}}
\def\Op{{\hat O}}
\def\id{{\hat I}}
\def\x{{\hat x}}
\def\P{{\hat P}}
\def\Px{\proj_x}
\def\Pr{\proj_{R}}
\def\Pl{\proj_{L}}
\def\ODR{O_{_{\rm DR}}(t)}
\def\ODRn{O_{_{\rm DR}}(n)}
\newcommand{\equa}[1]{Eq.~(\ref{#1})}


\title{Semiclassical approach to fidelity amplitude}

\author{Ignacio Garc\'{\i}a-Mata}
\affiliation{Instituto de Investigaciones F\'isicas de Mar del Plata (IFIMAR), CONICET--UNMdP,
Funes 3350, B7602AYL
Mar del Plata, Argentina.}
\affiliation{Consejo Nacional de Investigaciones Cient\'ificas y Tecnol\'ogicas (CONICET), Argentina}                         
\author{Ra\'ul O. Vallejos}
\email{vallejos@cbpf.br}
\homepage{http://www.cbpf.br/~vallejos}
\affiliation{ \mbox{Centro Brasileiro de Pesquisas F\'{\i}sicas (CBPF), 
              Rua Dr.~Xavier Sigaud 150, 
              22290-180 Rio de Janeiro, Brazil}}
              
\author{Diego Wisniacki}
\email{wisniacki@df.uba.ar}
\affiliation{\mbox{Departamento de F\'{\i}sica ``J. J. Giambiagi", 
             FCEN, Universidad de Buenos Aires, 1428 Buenos Aires, Argentina}}
\date{\today}

\begin{abstract}
The fidelity amplitude is a quantity  of paramount 
importance in echo type experiments.
We use semiclassical theory to study the average fidelity amplitude for quantum chaotic systems under external perturbation. 
We explain analytically two extreme cases: 
the random dynamics limit --attained approximately by strongly chaotic systems-- 
and the random perturbation limit, which shows a Lyapunov decay. 
Numerical simulations help us bridge the gap between both extreme cases.
\end{abstract}

\maketitle

\section{Introduction}
\label{sec1}
Irreversibility in quantum  mechanics cannot be explained in the 
same way as in classical mechanics. 
Instead of measuring how difficult it is to invert the individual 
trajectories, Peres proposed \cite{Peres1984} as a way to understand
quantum irreversibility to measure the difficulty to invert the dynamics. 
To put it in other words, to measure how difficult it is to implement 
a certain Hamiltonian or 
to \emph{know} the specifics of the dynamics, given that complete 
isolation is impossible to attain. 
As a consequence the fidelity --also known as Loschmidt echo (LE) 
\cite{Jalabert2001}--  was proposed as a measure of instability of a 
given  Hamiltonian under external perturbations. 
The LE is commonly defined as
\begin{equation}
M(t)= |O(t)|^2 \, ,
\end{equation}
where 
\begin{equation}
O(t)=\bra{\psi_0}e^{i H_\varepsilon t/\hbar}e^{-iH_0 t /\hbar}\ket{\psi_0}
\end{equation}
is the fidelity amplitude (FA) and 
$H_0$ and $H_\varepsilon$ differ by a perturbation which is usually taken 
to be an additive term like $\varepsilon V(q,p)$, so $\varepsilon$ is the 
perturbation strength.
The importance of the LE lies not only in that it can be used to 
identify quantum chaos but also in that it is a measurable 
quantity. 

There is a vast amount of work attempting to characterize
the decay regimes of the LE as universal \cite{Jalabert2001,Jacquod2001,Gorin2006,Jacquod2009}. 
After a short time transient, chaotic systems decay exponentially. For small 
perturbation strength -- Fermi Golden Rule regime -- the decay rate depends quadratically on the perturbation strength. 
For larger perturbation the decay rate is expected to be perturbation independent, and given by the largest Lyapunov exponent of the corresponding classical system. For regular systems the behavior is fundamentally different: for small perturbation strength the decay is Gaussian -- which can in fact lead to faster decay than chaotic systems \cite{Prosen2002,Prosen2002_2}-- 
and for larger values of the perturbation the decay is power law (see reviews  \cite{Gorin2006,Jacquod2009}).
However,  for chaotic systems, recent works  
\cite{Wang2004,Andersen2006,Pozzo2007,Natalia2009,Casabone2010}
have reported non-universal
oscillatory behavior in the decay rate of the LE as a function of the 
perturbation strength. 
In the way to elucidating the origin of such ``anomalous" behavior of 
the LE one key quantity to understand is the FA. 
The FA is important in itself for a number of reasons, 
the most important being that
in some of the echo experiments it is in fact the quantity that is measured \cite{Lobkis2003,Schafer2005,Schafer2005_2,Gorin2006_2,Lobkis2008,Kober2010}.

It is important to remark that in previous works the characterization of decay regimes
is done on the average behavior of the LE. Unlike most papers in the field -- with the exception of
\cite{vanicek2004} --
in this work we concentrate on the average FA. The fundamental difference with
\cite{vanicek2004}
 is that we obtain the decay regimes directly of the average FA, not requiring 
 to compute the average fidelity.
In order to do so we use the semiclassical theory known as dephasing 
representation (DR) \cite{vanicek2003, vanicek2004, vanicek2006}. 
We consider two limiting 
situations and obtain rigorous analytical expressions for the 
decay rate in both cases. 
In the first case we suppose
the dynamics is completely random.  
For maps on the torus each step is given 
by drawing two random numbers from a uniform distribution. This corresponds
to the limit of  infinite Lyapunov exponent. 
In this limit we obtain an expression for the decay of the FA that is valid 
for all times --
up to the saturation point. In fact it should be 
observed for very short times for any map -- our expression is exact for $t=1$. 
As the Lyapunov exponent is increased the random dynamics decay regime is observed for 
increasingly longer times.
We have already presented related results 
in \cite{nacho2011}. 
The other limiting case 
presupposes that the perturbation is completely random. 
That is, after each step of the map the perturbation contributes with a random phase 
to each trajectory.
It shall be seen that this assumption can be justified in the large perturbation strength limit because semiclassical
calculations involve complex exponentials of such strengths.
In that case  --using the DR and transfer matrix theory-- we show that the 
asymptotic decay rate of the FA is 
controlled by the largest classical 
Lyapunov exponent $\lambda$.  In other words, 
we are capable of predicting , analytically, the appearance of a Lyapunov regime in the
average FA rather that the fidelity itself (i.e. the square of the modulus of the FA).

Even though in a realistic scenario one may not see clearly the
appearance of  the limiting behaviours, traces of one (or both)
of them are usually present.
For short times, on average the decay rate is 
that predicted for random dynamics. 
This result is valid for longer times as the dynamics becomes 
effectively more random --i.e. as $\lambda$ grows. 
Then, for large perturbations, there is a crossover to an asymptotic
regime where the decay rate is $\lambda/2$, independently of the 
perturbation.
The crossover time is strongly perturbation dependent.
We 
exhibit numerical simulations introducing a perturbation model 
where the amount of randomness can be varied. 
In this model the decay of the FA is well described by a sum of 
two terms, each one corresponding to one of the limiting behaviors. 
The relative weight of the terms depend both on $\lambda$ and 
on the perturbation (amplitude, length scale).

The paper is organized as follows. 
Section~\ref{sec2} gives a brief introduction to the dephasing 
representation (DR). 
This semiclassical approximation is the basis of two analytically
tractable models to be developed in the subsequent sections.
The random-dynamics approach, which is valid for large Lyapunov
exponents, is described in Sect.~\ref{sec3}. 
Using the baker-map family as a model of chaotic dynamics, and 
employing the transfer matrix method, we argue in Sec.~\ref{sec4} 
that for large perturbations the asymptotic decay is ruled by the 
Lyapunov exponent.

Section~\ref{sec5} presents a brief 
study of mixed regimes, 
where the decay of the FA is bi-exponential. 
We show in some numerical examples that the decay rates are still 
given by the models mentioned above. 
An empirical expression for the crossover time is found. 

We conclude in Sec.~\ref{sec6} with a summary of our results 
and some final remarks.
\section{Semiclassical dephasing representation of 
the fidelity amplitude}
\label{sec2}
Recently \cite{vanicek2003, vanicek2004, vanicek2006} 
the dephasing representation was introduced to give a compact and 
efficient way to compute semiclassically the FA. 
The derivation of the DR starts by replacing the quantum propagators 
by the semiclassical Van Vleck propagator --this is the standard 
semiclassical approach. 
The first innovation is to use the initial value representation 
\cite{Miller1970,Miller2001} 
for the Van Vleck propagator. 
In this way a full semiclassical --so-called uniform-- expression for 
$O(t)$ can be obtained. 
However, this expression is still too difficult to be computed and a further 
approximation is needed which involves using trajectories of $H_\varepsilon$ 
with slightly different initial conditions but which remain close to the 
trajectories of $H_0$ up to a certain time.
The validity of this ``dephasing trajectories'' argument was justified 
using the shadowing theorem \cite{vanicek2004}. 
One of the forms of 
the FA obtained using the DR looks like 
\begin{equation}
	\label{eq:odr}
\ODR=\int dq dp W_{\psi}(q,p) \exp(-i \Delta S_\varepsilon(q,p,t)/\hbar)
\end{equation}
where $W(q,p)$ is the Wigner function of the initial state $\psi$ and
\begin{equation}
\Delta S_\varepsilon(q,p,t)=-\varepsilon\int_0^t d\tau V(q(\tau),p(\tau)) 
\end{equation}
is the action difference evaluated along the unperturbed classical trajectory. 
In this way the decay can be attributed to the dephasing produced 
by the perturbation of the actions --thus the name DR.
However numerics show that \equa{eq:odr} also accounts for decay due to 
classical overlaps \cite{vanicek2006}.

The DR is a useful tool to compute fidelity and to assess fidelity decay.  
Actually, when referring to fidelity decay \emph{regimes} one refers to an 
average behavior.
Throughout this work by average we mean average over initial states. 
The average FA over
a number $n_r$ of initial states is
\begin{equation}
\overline{\ODR}=
\frac{1}{n_r}\sum_{i=1}^{n_r} \int dq dp W_{\psi_i}(q,p) 
\exp(-i \Delta S_\varepsilon(q,p,t)/\hbar),
\end{equation}
If in particular, $\{\psi_i\}_{i=0}^{n_r}$ is some complete set, 
the Wigner function of an incoherent sum of such a set is a constant. 
We then get
\begin{equation}
	\label{eq:afa}
\overline{\ODR}=
\frac{1}{\cal V}\int dq dp \exp(-i \Delta S_\varepsilon(q,p,t)/\hbar),
\end{equation}
where ${\cal V}$ is the volume of phase space. 
 This expression was used in \cite{vanicek2004arxiv} to compute the 
 FA averaged over initial states. 
Henceforth we take ${\cal V}=1$. 

Equation~(\ref{eq:afa}) is the starting point of our studies.
By making some additional assumptions on the dynamics and/or 
the perturbation we shall be able to derive analytical expressions 
for the decay rate of the average FA (next sections).
Being at the basis of our future developments, it is necessary
to be sure that (\ref{eq:afa}) gives indeed an accurate description
of the FA decay. 
 The DR has successfully passed many tests in the last 
years 
 \cite{Wang2004,wang2005,vanicek2006}, especially in quantum chemistry 
\cite{Li2009,Zimm2010,Zimm2010_2,Wehrle2011}. There has been a similar approach 
to the dephasing representation in the case of linear response functions for electronic spectra \cite{Shi2005}. 
We want, though, to verify its performance in the
specific systems that shall be used in this paper. 

For the numerical comparisons we have preferred to use quantum 
maps on the torus. These maps possess all the 
essential ingredients of the chaotic dynamics and are, at the same time,
extremely simple from a numerical point view --both at the classical
and at the quantum levels.  
The periodic boundary conditions that the torus geometry imposes 
translates in a discretization of both $q$ and $p$ upon quantization.
The Hilbert space is then finite dimensional (dimension $N$) 
corresponding to an adimensional Planck's constant 
$\hbar=1/(2\pi N)$ (we assume the torus has unit area). 
Position and momentum basis are related by the discrete Fourier 
transform (DFT). 
In such a setting the quantum map corresponds to a unitary 
operator $U$ \cite{haake}.
In particular we choose quantum maps that can be represented 
in the split-operator form:
\begin{equation}
\label{eq:qmap}
U=e^{ -i 2\pi N T(\hat{p})}e^{-i 2\pi N V(\hat{q})}.
\end{equation}
The reason of using maps with this structure is twofold: 
(i) Many well known classical maps fall in this category, e.g., 
the standard map, 
the kicked Harper map, 
the sawtooth map, 
as well as some cat maps \cite{haake,ozorio}.
(ii) They allow a very efficient numerical implementation 
due to the possibility of using the discrete Fourier transform.

The classical version of $U$ is
\begin{equation}
\left.
\begin{array}{ccl}
\bar{p}&=&p-\frac{dV(q)}{dq}\\
\bar{q}&=&q +\frac{dT(\bar{p})}{d\bar{p}}
\end{array}
\right\} \quad ({\rm mod}\  1).
\end{equation}
Here we shall consider the perturbed cat maps given by 
\cite{dematos}
\begin{equation}
\left.
\begin{array}{ccl}
\label{eq:pcat}
\bar{p}&=&p-a\,q+K f(q)\\
\bar{q}&=&q-b \bar{p}+\tilde{K} h(\bar{p})
\end{array}
\right\}\quad ({\rm mod}\  1).
\end{equation}
In particular, for the numerics in this section and in Sec.~\ref{sec3} 
we consider the perturbing ``forces''
\begin{eqnarray}
\label{eq:pert}
f(q)&=&2 \pi \left[ \cos(2 \pi q) - \cos(4 \pi q) \right] \, , \\ 
h(\bar{p}) &=&0 \, .
\end{eqnarray}
In Eq.~(\ref{eq:pcat}) we set $K\ll 1$ and $a,b$ such that the 
resulting map is completely chaotic (hyperbolic). 
In such a case the Lyapunov exponent is approximately
\begin{equation}
\lambda \approx \log \frac{1 + ab + \sqrt{ab(4+ab)}}{2} \, .
\end{equation}

Time is discrete in maps, so from now on we use the integer $n$ 
to count time steps.
We shall use the $K$-dependence of the quantum map $U$ to define
the fidelity amplitude as
\begin{equation}
\label{eq:odt}
O(n)=\bra{\psi_0}((U_{K'}^\dag)^n U_{K}^n\ket{\psi_0}  \, .
\end{equation}
Accordingly, the perturbation strength is given by
\begin{equation}
\varepsilon=|K-K'|   \, .
\end{equation}

In order to calculate the semiclassical DR expression for the 
average FA 
we define a $n_s=n_q \times n_p$ grid of initial positions and momenta. 
Thus we get
\begin{equation}
\label{eq:ODRn}
\overline{\ODRn}=
\frac{1}{\Omega}\sum_{q,p}\exp[-i 2 \pi N \Delta S_\varepsilon(q,p,n)].
\end{equation}
where $q=i/n_q$, $p=j/n_p$ 
(with \mbox{$i=0,\ldots,n_q-1$}, \mbox{$j=0,\ldots,n_p-1$}), 
and  
$\Omega=1/n_q n_p$.
The action difference is  
\begin{equation}
\label{eq:discrac}
\Delta S_\varepsilon(q,p,n)=-\varepsilon\sum_{i=1}^{n}F(q_i)
\end{equation}
with $F(q)=\int f(q) dq$
and the points $(q_i,p_i)$ are the successive iterates of the 
classical map of \equa{eq:pcat}. 

\begin{figure}
\begin{center}
\includegraphics[width=0.9\linewidth]{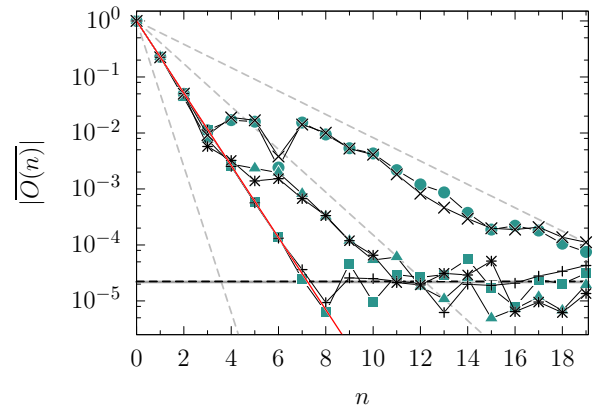} 
\caption{(color online) 
Average fidelity amplitude as a function of
time 
for the perturbed cat map of \equa{eq:pcat} 
for 
$a=b=1$  ($\lambda=0.96$; {\color{verde2}{\large $\bullet$}} and $\times$), 
$a=b=2$  ($\lambda=1.76$; {\color{verde2}{\large $\blacktriangle$}} and $*$)
and 
$a=b=20$ ($\lambda=5.99$; {\color{verde2}$\blacksquare$}  and $+$).  
Both $\times$, $*$, and $+$ symbols correspond to the semiclassical
approximation (DR) while {\color{verde2}{\large $\bullet$}}, 
{\color{verde2} {\large $\blacktriangle$}}  and 
{\color{verde2}$\blacksquare$} 
correspond to full quantum evolution. 
The three dashed (grey) lines represent the exponential decay 
$\exp(-\lambda t/2)$. 
The solid (red) line is
$\exp(-\Gamma t)$ with $\Gamma$
computed semiclassically (see Sec.~\ref{sec3}). 
We set
$\varepsilon/\hbar=6.433$ which corresponds to  
$\Gamma=1.484$.
The horizontal lines indicate saturation values: 
dashed (black) line is $1/\sqrt{n_s}$, with $n_s=2\times10^9$ (DR); 
solid (grey) line is 
$1/\sqrt{N\times n_r}$ with $N=2^{18}$ (Hilbert space dimension) 
and $n_r=8000$ (number of initial states).  
\label{fig:comp}}
\end{center}
\end{figure}

In figure~\ref{fig:comp} we show the time evolution of the fidelity 
amplitude, comparing the semiclassical approximation (DR) with the 
full quantum evolution. 
Three perturbed cat maps were considered (with Lyapunov exponents
$\lambda=0.96,\, 1.76,\, 5.99$) under the same perturbation.
For the full quantum calculation we used a dimension $N=2^{18}$ and 
averaged over $n_r=8000$ initial coherent states. 
For the DR simulation the integral (\ref{eq:afa}) was discretized,
the total number of 
initial conditions being $n_s=2\times10^9$.

We see that the agreement between quantum and semiclassical 
calculations is quite good in both cases. 
This makes us confident on the dephasing representation as
a starting point for the analytical understanding of the 
average FA.
In the next sections we shall describe two models that 
explain satisfactorily the main features observed in 
figure~\ref{fig:comp}: for a large Lyapunov exponent the
dynamics can be considered as being essentially random and 
the decay is single-exponential with a rate that depends 
only on the perturbation (Sect.~\ref{sec3}).  
If the Lyapunov exponent is small and/or the perturbation
is large enough the asymptotic decay rate is independent 
of the perturbation, equal to half the Lyapunov exponent, 
(Sect.~\ref{sec4}).

To conclude this section we briefly address the matter of 
saturation values.
The average of $n_s$, modulo one, complex numbers with 
uniform random phases goes like $1/\sqrt{n_s}$. 
Therefore for chaotic systems and long enough times 
the DR yields approximately
$\overline{\ODR}\sim 1/\sqrt{n_s}$. 
On the other hand,
it is well known \cite{gutierrez2009} that 
the saturation value of \equa{eq:odt} is determined 
by $1/\sqrt{N}$, with $N$
the size of the Hilbert space.
Therefore in the full quantum simulation the saturation 
value is $\sqrt{N\times n_r}$. 
Notice that in the simulations 
we chose $n_s,n_r,N$ 
so that the saturation values of both quantum and semiclassical 
calculations approximately coincide. 
%
\section{Fidelity amplitude for random dynamics}
\label{sec3}
Recently \cite{Goussev2008} the DR was used to derive analytically 
the decay rate of the FA for perturbations acting on a small
portion of phase space. 
The advantage of using local perturbations
is that we can suppose that 
the trajectory becomes uncorrelated between successive hits 
to the perturbed region. 
It can be argued that the same effect can be achieved for 
non-local perturbations provided that $\lambda$ is very 
large a \cite{nacho2011}.

\begin{figure}
\begin{center}
\includegraphics[width=0.9\linewidth]{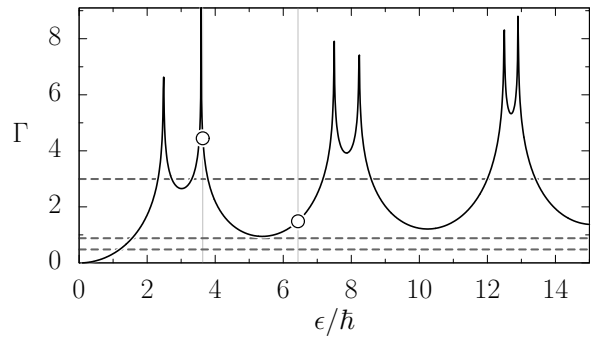} 
\caption{(color online) 
Decay rate $\Gamma$ computed from Eqs.~(\ref{eq:gamma1}) and 
(\ref{eq:gamma2}) as a function of the rescaled perturbation. 
The points indicate the values chosen in 
figure~\ref{fig:comp}  ($\varepsilon/\hbar=6.433$) and 
figure~\ref{fig:comp2} ($\varepsilon/\hbar=3.635$). The dashed horizontal lines
are the corresponding $\lambda/2$ used in figure 1 ($0.96/2$, $1.76/2$ and $5.99/2$).
\label{fig:rnddyn}}
\end{center}
\end{figure}

\begin{figure}
\begin{center}
\includegraphics[width=0.9\linewidth]{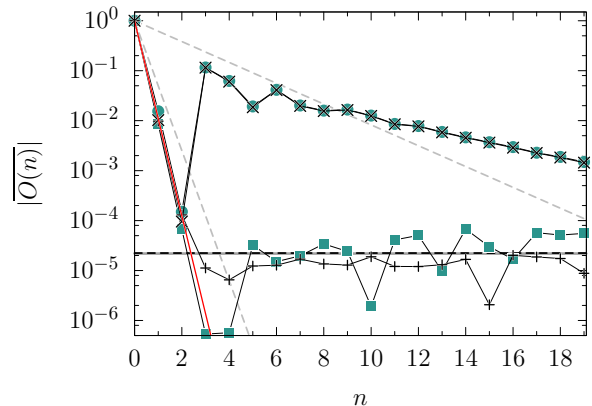} 
\caption{(color online) 
Same as figure~\ref{fig:comp} with $\varepsilon/\hbar=3.635$ corresponding
to $\Gamma=4.448$. 
\label{fig:comp2}}
\end{center}
\end{figure}

In this section we study the behavior of $\overline{O_{DR}}(n)$
in the $\lambda\to\infty$ limit by assuming that 
the dynamics is purely random.
This evolution is completely stochastic in the sense that 
there is no correlation for the different times of the evolution.
To compute $\overline{O_{DR}}(n)$, we make a partition of the 
phase space in $N_c$ cells. 
We consider that the probability to jump from cell to any other 
in phase space is uniform.
Therefore it is straightforward to show that the mean FA results 
\bea
\overline{O_{DR}}(n) &=&\frac{1}{N_c^n}
\sum_{j_1}...\sum_{j_n} 
\exp [-i (\Delta S_{\varepsilon,j_1} +...+ \Delta S_{\varepsilon,j_n})/\hbar ] \nonumber \\ 
& = &  
\left[\frac{1}{N_c} \sum_{j} \exp \left( -i \Delta S_{\varepsilon,j} /\hbar \right) \right]^{n}  \, ,
\eea
where $\Delta S_{\varepsilon,j_k}$ is the action difference evaluated 
in the cell $j_k$ at time $k$.
The continuous limit is approached when $N_c \rightarrow \infty$. 
So,  $\overline{O_{DR}}(n)$ results
\be
\overline{O_{DR}}(n)=
\left( \int \exp \left[ -i \Delta S_\varepsilon(q,p) /\hbar \right] dq dp\right)^n \, ,
\ee
where $\Delta S_\varepsilon(q,p)$ is the action difference after one step 
of the map.
This exponential decay can be rewritten as
\be
\label{eq:gamma1}
\overline{O_{DR}}(n)=\exp (-\Gamma \, n)  \, ,
\ee
with
\be
\label{eq:gamma2}
\Gamma=-\log \Big|\int \exp [ -i  \Delta S_\varepsilon(q,p)/\hbar] dq dp\Big| \, .
\ee
 Equation (\ref{eq:gamma2}) is one of the key results of this work: 
if the dynamics is (uncorrelated) random then the average fidelity amplitude decays exponentially with 
decay rate $\Gamma$.
In figure~\ref{fig:rnddyn} we plot $\Gamma$ as a function of the rescaled 
perturbation strength $\varepsilon/\hbar$ for the map of \equa{eq:pcat}. 
The circles mark the perturbation strength values used in 
Figs.~\ref{fig:comp} and \ref{fig:comp2}. 
%
Both in Figs.~\ref{fig:comp} and \ref{fig:comp2} the agreement with
the random-dynamics approach 
is well observed for initial times. 
If $\lambda$ is large enough 
($\lambda=5.99$ for {\color{verde2}$\blacksquare$} and $+$ symbols in Figs.~\ref{fig:comp} 
and \ref{fig:comp2}), 
the dynamics can be considered as random for 
practical purposes 
and the decay rate --up to saturation-- 
is given by $\Gamma$.
%
%
In Sec.~\ref{sec4} we shall see in a specific deterministic model 
(the baker family) how a large Lyapunov exponent leads to an effective 
decorrelation of the successive points in a trajectory, and therefore
the map behaves as if it were random.

We point out that in Figs.~\ref{fig:comp} and \ref{fig:comp2}, 
after the inital
decay following the random dynamics predictions, there is a crossover 
to a different decay rate.  
In figure~\ref{fig:comp} the second decay rate is $\lambda/2$ 
while in figure~\ref{fig:comp2} it is not. 
In Secs.~\ref{sec4} and \ref{sec5} we analyze in more detail this crossover behavior.
\section{The random-perturbation limit}
\label{sec4}
In the previous section we showed that a large Lyapunov exponent 
leads to a decay that is well described by the random dynamical 
model.
Here we argue that for large enough perturbations a 
Lyapunov decay must be expected.

Key to our analysis is the implementation of the transfer matrix
method to calculate the semiclassical average fidelity 
amplitude.
The transfer matrix method is a standard tool in Statistical
Mechanics for the calculation of the partition function 
of Ising lattices \cite{dobson69,borzi87,reichl}. 
Due to the analogy between Ising partition functions (for
imaginary temperature) and semiclassical periodic orbit sums,
the method has also found applications in the Quantum Chaos
domain \cite{gutzwiller,kaplan96,smilansky03,carlo10}.

We start by analysing the particularly simple case of the 
baker map and a random perturbation. 
Then, we provide analytical and numerical evidence showing 
that such a Lyapunov decay should also be observed in a 
larger class of chaotic maps submitted to nonrandom (large enough) 
perturbations.
\subsection{The baker map and the transfer matrix method}
The baker's transformation is an area preserving,
piecewise-linear map of the unit square defined by \cite{arnold}
\bea
q_1 &=& 2 q_0 - \mu_0    \,, \\
p_1 &=&  (p_0 + \mu_0)/2 \,,
\eea
where $\mu_0=[2q_0]$, the integer part of $2q_0$.
This map admits a very useful description in terms of a 
complete symbolic dynamics \cite{adler98}. 
A one to one correspondence between phase space coordinates 
and  binary sequences,
\be
(p,q) \leftrightarrow 
\ldots 
\mu_{-2} 
\mu_{-1}
\cdot
\mu_{0} 
\mu_{1} 
\mu_{2} 
\ldots
~~~,~ \mu_i=0,1 ~,
\label{symbol}
\ee
can be constructed in such a way that the action of the map is
conjugated to a shift map.  
The symbols are assigned as follows:
$\mu_i$ is set to zero (one) when the $i$--th iteration of 
$(p,q)$ falls to the left (right) of the line $q=1/2$, i.e. 
$[2q_i]=\mu_i$. 
Reciprocally, given an itinerary
$\ldots 
\mu_{-2} 
\mu_{-1}
\cdot
\mu_{0} 
\mu_{1} 
\mu_{2} 
\ldots$, 
the related phase point is obtained through the especially
simple binary expansions
\bea
q & = & \sum_{i=0}^{\infty} \frac{\mu_i   }{2^{i+1}} \equiv 
                                                 .\mu_{0}  \mu_{1} \ldots  \, , \\
p & = & \sum_{i=1}^{\infty} \frac{\mu_{-i}}{2^i    } \equiv 
                                                 .\mu_{-1} \mu_{-2} \ldots \, .
\label{pqsym}
\eea

For simplicity, let us initially consider a perturbation that
depends only on $q$. 
Using the binary expansions defined above, the DR average fidelity
amplitude (\ref{eq:ODRn},\ref{eq:discrac}) reads:
\bea
& & \overline{\ODRn}= \int_0^1 dq_0 \, \nonumber \\
& &  
e^{i \varepsilon \left[ 
F( 
.\mu_{0} 
\mu_{1} 
\mu_{2} 
\ldots) + 
F( 
.\mu_{1} 
\mu_{2} 
\mu_{3} 
\ldots) + 
\ldots  +
F( 
.\mu_{n-1} 
\mu_{n} 
\mu_{n+1} 
\ldots) \right] }  \, .
\label{intbaker}
\eea
Here and in the rest of this section, for the sake of a lighter notation,
we write just $\varepsilon$ in place of $\varepsilon/\hbar$.

In order to introduce the transfer matrix method it is necessary
to truncate the binary expansions at a finite length $L$, i.e., 
\be
q \approx \sum_{i=0}^{L-1} \frac{\mu_i}{2^{i+1}} \, .
\ee
Equivalently one may think that we are treating exactly a 
perturbation which is constant over vertical strips of equal 
width $1/2^L$ (we adopt this point of view).
With this proviso the integral (\ref{intbaker}) becomes a finite 
sum: 
\bea
& & 
\overline{\ODRn}=2^{-(n+L-1)} \sum_{\mu_{0}, \ldots, \mu_{n+L-2}=0,1} \nonumber \\
& & 
e^{i \varepsilon 
\left[ 
F(. \mu_{0} \ldots \mu_{L-1})+ 
F(. \mu_{1} \ldots \mu_{L  })+
\ldots+
F(. \mu_{n-1} \ldots \mu_{n+L-2}) 
\right] }  \, .
\eea
The prefactor $2^{-(n+L-1)}$ represents the {\em area} (weight, probability)
of the region in phase space corresponding to the finite symbolic string
$\mu_{0}, \ldots, \mu_{n+L-2}$ (the area depends on the length 
only, not on the particular code). 
These regions taken together form a partition of phase space
into disjoint elements.
All the trajectories starting in a given element have the same action.

\begin{figure}[htp]
\begin{center}
\includegraphics[angle=0, width=0.9\linewidth]{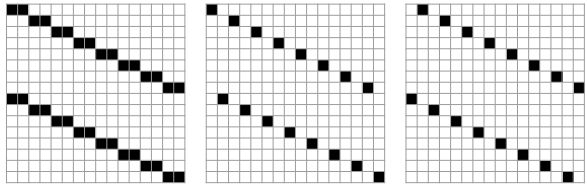} 
\caption{The transfer matrix $M$ (left) can be split into two
unitary matrices $U_1$, $U_2$ (center, right). 
Nonzero elements are shown as black squares ($L=5$, schematic). }
\label{fig41}
\end{center}
\end{figure}

The trick now is to express the sum above as a product of $n$ matrices.
Define the indices
\bea
   k_0 & = & 2^{L-1} \times  .\mu_{0} \ldots \mu_{L-2}   \, , \label{k0} \\
   k_1 & = & 2^{L-1} \times  .\mu_{1} \ldots \mu_{L-1}   \, , \label{k1} \\
\ldots & = & \ldots                           \, , \\
   k_n & = & 2^{L-1} \times  .\mu_{n} \ldots \mu_{L+n-2} \, .
\eea
Then we can write
\be
\overline{\ODRn} = 2^{-(n/2+L-1)} 
\sum_{k_0, \ldots, k_n}
M_{k_0,k_1} \, M_{k_1,k_2} \ldots M_{k_{n-1},k_n}  \, ,
\label{prodMMM}
\ee
where the nonzero elements of the matrix $M$ are given by
\be
M_{k_0,k_1} = 
\frac{1}{\sqrt{2}} \, e^{i \varepsilon F(. \mu_{0} \ldots \mu_{L-1})} 
\label{Mk0k1}
\ee
(the convenience of having introduced a factor $\sqrt{2}$ will 
become clear soon).
It is implicit in this definition that $k_0$ and $k_1$ must satisfy
the {\em shift condition}, i.e., they must share the bits 
$\mu_{1} \ldots \mu_{L-2}$
[see Eqs.~(\ref{k0},\ref{k1})], otherwise the matrix
element is set to zero.
The shift condition determines that nonzero elements
are located on two staircases of slope 1/2 (``baker structure''), 
explicitly given by
\be
k_1 = 2 k_0 - 2^{L-1} \mu_0 +  \mu_{L-2} 
\ee
(see figure~\ref{fig41}).

If one defines the unit-norm vector
\be
| 1 \rangle = \frac{1}{\sqrt{2^{L-1}}} (1,1,\ldots,1) \, ,
\ee
then Eq.~(\ref{prodMMM}) can be written in the compact way
\be
\overline{\ODRn} = 2^{-n/2} \, \langle 1 | \, M^n \, | 1 \rangle  \, .
\label{Iofn}
\ee

This is a remarkable result: the decay properties of the
average fidelity amplitude are embodied in the spectrum
of a finite matrix.
In particular, the asymptotic decay is ruled by the 
largest eigenvalue (in modulus) of $M$.
The special staircase structure of the transfer matrix 
allows us to deduce some general properties of its
spectrum.

For a fine discretization, the phase in Eq.~(\ref{Mk0k1}) 
varies by small jumps along the staircases of $M$, the size 
of the jumps depending both on the characteristic spatial 
scale of $F(q)$ (``correlation length'') and on the 
perturbation amplitude $\varepsilon$. 
As the correlation length decreases and/or the amplitude 
grows, neighboring matrix elements become increasingly 
decorrelated. 
This leads us to consider, as a reference, an ensemble 
of random transfer matrices having the same structure 
as $M$, but whose phases are i.i.d. random variables chosen 
from a uniform distribution in $[0,2\pi]$. 

Random or not, because of its particular structure, 
the matrix $M$ can be split as a sum of two 
unitary matrices $U_1,U_2$ (see figure~\ref{fig41}):
\be
M=\frac{1}{\sqrt 2} \left( U_1 + U_2 \right) \,.
\ee
When $U_1$ and $U_2$ are random unitary matrices drawn 
independently from the Circular Unitary Ensemble (CUE)
\cite{mehta,haake}, 
analytical and numerical studies say that the spectrum of 
$M$ lies almost entirely inside the unit circle in the 
complex plane. 
As the dimension of $U_i$ is increased, the largest 
eigenvalue (in modulus) tends to the unit circle.
In the limit of infinite dimensional matrices, all the 
spectrum is contained in the unit disk 
\cite{gorlich,carlo10}.  

\begin{figure}[htp]
\begin{center}
\includegraphics[angle=0, width=0.95\linewidth]{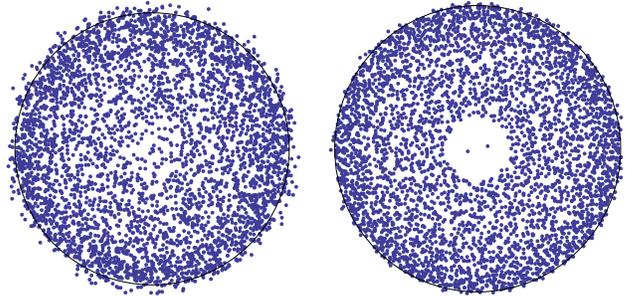} 
\caption{(Color online) 
Eigenvalues $z_i$ of random matrices having the baker
structure. Shown are the combined spectra of 64/8 matrices 
of dimension 64$\times$64/512$\times$512 (left/right). 
Circles of unit radii were drawn for reference.}
\label{fig42}
\end{center}
\end{figure}

We have not attempted an analytical study of the spectral 
properties of random matrices with the baker structure. 
However, our numerical calculations show that they 
possess very similar statistical properties to those of 
the normalized sum of two CUE matrices 
(as far as the spectrum close to the unit circle is concerned).
Figure~\ref{fig42} shows clearly that most eigenvalues 
are distributed inside the unit circle, the relative number 
of ``outsiders'' decreasing as dimension in increased.

So, as the leading eigenvalues of $M$ are close to the 
unit circle, we conclude that for a random perturbation
the asymptotic decay of $\overline{\ODRn}$ is governed by 
the prefactor in (\ref{Iofn}), thus we have
\be
|\overline{\ODRn}|^2 \sim  2^{-n}   \, .
\ee
This is an exponential decay, with a rate given by the Lyapunov
exponent of the baker map.  This is the other key result of the present work:
starting from the FA, and using the DR and the transfer matrix method 
we obtain that the asymptotic decay of the average FA 
is given by $\lambda/2$ 
(consequently the fidelity decays with the rate lambda).

\begin{figure}[htp]
\begin{center}
\includegraphics[angle=0, width=0.9\linewidth]{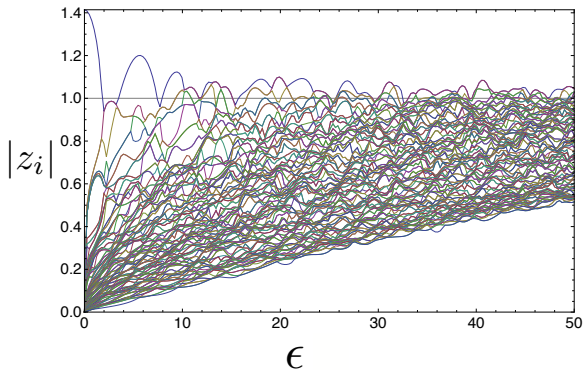} 
\caption{(Color online) Transfer matrix eigenvalues (moduli) as 
a function of perturbation amplitude. 
The perturbation $F(q)= \sin(2 \pi q)-\sin(4 \pi q)/2$ 
was discretized in a lattice of 256 points ($L=7$).}
\label{fig43}
\end{center}
\end{figure}

After having analysed the random transfer-matrix model, 
we question to what extent a nonrandom perturbation may lead 
to a random-matrix behavior.
As mentioned before, e.g., a smooth but large perturbation 
may cause a randomization of the phases (\ref{Mk0k1}).
In order to check this expectation we consider now
the baker dynamics with the action difference of 
Eqs.~(\ref{eq:discrac},\ref{eq:pert}), i.e., 
$F(q)= \sin(2 \pi q)-\sin(4 \pi q)/2$.
The spectrum of the corresponding $M(\varepsilon)$ 
is shown in figure~\ref{fig43}.
There we see that for, say, $\varepsilon > 10$ the leading
eigenvalues settle down in a small neighborhood of
the unit circle.
Then, for such large perturbations $|\overline{\ODRn}|^2$ decays with 
a rate which is close to the Lyapunov exponent of the 
classical map.

Note, however, that in typical numerical simulations the decay 
rate may exhibit some departure from the Lyapunov exponent. 
This may be due to two reasons. 
First, the perturbation amplitude may be not large enough for
the settling down of the Lyapunov regime.  
Second, even for completely random perturbations, if $M$ is
finite-dimensional then the leading eigenvalue will be in 
general outside the unit circle. 
It is true that, as the dimension $N$ of $M$ is increased, the
leading eigenvalue tends to the unit circle, but it does so very
slowly, possibly like $N^{-1/3}$ \cite{carlo10}.
So, in practice, some finite deviations from the Lyapunov decay 
rate should not be ruled out.

\subsection{Extensions}
Here we present some extensions of the application of
the transfer matrix method to account for more general
maps and/or perturbations.

First note that perturbations that depend both on $q$ and $p$ 
can easily be accomodated in the transfer-matrix scheme, for, 
according to (\ref{symbol}), it requires only adding some symbols
for the discretization in $p$. 

Now consider the $B$-adic baker map defined by (\cite{sano00})
\bea
q_1 &=& B q_0 - \nu_0    \,, \\
p_1 &=&  (p_0 + \nu_0)/B \,,
\eea
where $B$ is an integer $\ge 2$, and $\nu=0,1,\ldots,B-1$.
This map is completely analogous to the usual ($B=2$) baker
map, but with expansion rate $B$. 
At the symbolic level it corresponds to a full shift on $B$ 
symbols \cite{adler98}. 
Its Lyapunov exponent is $\log B$.
Thus, by making $B$ large enough we shall be able to enter 
the random-dynamics regime (Sect.~\ref{sec3}).

The transfer matrix $M$ is constructed in a similar way as 
we did for the baker: basically, all we have to do is to 
change from base-2 to base-$B$ in the standard-baker formulas. 
By making the appropriate substitutions, and skipping the details, 
one arrives at the following expression for the average FA
\be
\overline{O_B(n)} = B^{-n/2} \, \langle 1 | \, M_B^n \, | 1 \rangle  \, ,
\label{Iofn2}
\ee
with $| 1 \rangle$ a unit-norm vector of length $B^{L-1}$.

For a perturbation that only depends on $q$ the non-null 
elements of $M_B$ read
\be
\left( M_B \right)_{k_0,k_1} = 
\frac{1}{\sqrt{B}} \, e^{i \varepsilon F( .\nu_{0} \ldots \nu_{L-1})} \, ,
\label{MBk0k1}
\ee
where $\nu_i$ are the ``decimal'' digits of the $B$-ary 
expansion of the coordinate $q$, truncated at depth $L$. 
The non-null elements of $M_B$ lie along $B$ flights of 
stairs of slope $1/B$ (riser=1, tread=$B$).
In analogy to the $B=2$ case, we can split  $M_B$ into 
a sum of unitary matrices:
\begin{equation}
M_B=\frac{1}{\sqrt{B}} \sum_{i=1}^B U_i	\; .
\end{equation}

\begin{figure}[htp]
\begin{center}
\includegraphics[angle=0, width=0.9\linewidth]{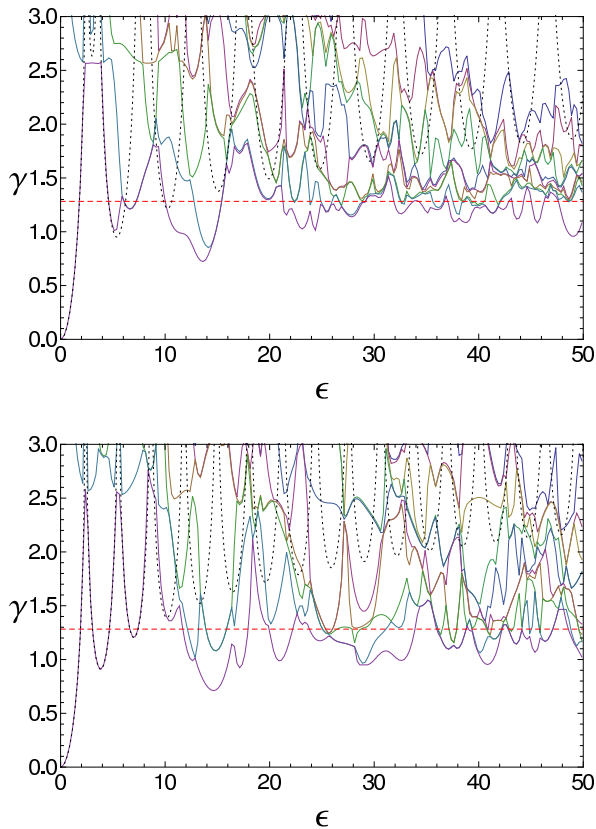} 
\caption{(Color online) 
Transfer matrix 
decay rates for
the $B=13$ baker as a function of perturbation amplitude. 
We considered two perturbed actions 
$F(q)= \sin(2 \pi q)-\sin(4 \pi q)/2$ (top panel) and
$F(q)= \sin(2 \pi q)$ (bottom panel).
These functions were discretized in a lattice of 2197 points ($L=3$).
Dashed horizontal lines correspond to Lyapunov decay, i.e., $(\log B)/2$.
Dotted lines are the predictions of the random-dynamics model.}
\label{fig44}
\end{center}
\end{figure}

Again, in the case that $U_i$ are large independent 
CUE matrices, it is possible to show that the spectrum 
of $M_B$ is almost completely contained in the unit disk 
in the complex plane \cite{gorlich}. 
The same behavior is expected for the transfer matrix
of the baker family, when a nonrandom but large
enough perturbation is considered.
In such a case, one would have the asymptotic decay
\be
|\overline{O_B(n)}| \sim  e^{-\lambda_B n/2}   \, ,
\ee
with $\lambda_B =\log B$, the Lyapunov exponent of the
classical $B$-baker map.
As an example, we show in figure~\ref{fig44} the transfer
matrix spectrum for $B=13$ and a smooth perturbation.
Plot are the decay rates 
\begin{equation}
\gamma_i=\frac{\log B}{2} -\log \left| z_i \right|   \, ,  
\end{equation}
where $\left\{z_i\right\}$ stand for the eigenvalues 
of $M_B$.

Focusing on the asympotic decay, we see that for small 
perturbation amplitudes the decay is very well described 
by random-dynamical model, as was expected 
from the discussion of Sect.~\ref{sec3}.
Note, however, that the extension of the random-dynamic 
regime is sensitive to the {\em form} of the perturbation.
For large $\varepsilon$ the leading decay rate fluctuates 
around the Lyapunov value $(\log B)/2$, in a similar way
as we had seen in the case of the standard baker.

The baker family permits a transparent estimation of the 
crossover point  between random-dynamics and Lyapunov regimes.
For a $q$-dependent perturbation and a $B$-baker map,
the average fidelity amplitude is given by
\be
\overline{\ODRn}= \int_0^1 dq_0 \, 
e^{i \varepsilon \left[ F(q_0) + F(q_1)+ \ldots + F(q_{n-1}) \right]} \, .
\ee
Here it is implicit that
$q_1,q_2,\ldots$ are related to $q_0$ by the dynamics.
Alternatively we can take the $q_i$ to be independent and write
\bea
& & \overline{\ODRn}= \int dq_0 \, dq_1 \, \ldots dq_{n-1}
e^{i \varepsilon \left[ F(q_0) + F(q_1)+ \ldots  F(q_{n-1}) \right]} 
\times \nonumber \\
& & 
P(q_0,q_1,\ldots,q_{n-1}) \, ,
\eea
where
\be
P(\ldots)=
\delta(q_1-g(q_0))  
\delta(q_2-g^2(q_0)) 
\ldots 
\delta(q_{n-1}-g^{n-1}(q_0)) 
\, ,
\ee
and $g(q)=Bq-[Bq]$.

\begin{figure}[htp]
\begin{center} 
\includegraphics[angle=0, width=0.95\linewidth]{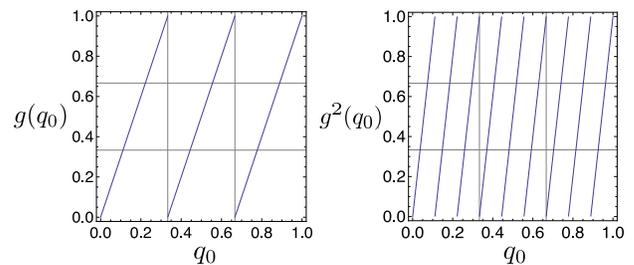} 
\caption{The probability distribution is a product of delta
functions $\delta(q_1-g(q_0)) \delta(q_2-g^2(q_0)) \ldots$. 
Gridlines define the critical square binning.
All bins contain the same probability, meaning that the
histogram is constant, therefore separable.
The same is true for any bin size larger than $1/B$ 
($B=3$ in these plots). }
\label{fig45}
\end{center}
\end{figure}

If the joint probability distribution $P$ were factorable,
then we would recover the formula for random dynamics.
Strictly speaking this factorization is impossible for
a deterministic dynamics,
however, when considering $P$ in the context of the integral 
above one sees that $P$ admits some coarse graining without
affecting the integral. 
This is true whenever the scale for the smoothing is
smaller than the typical scale of variation of the phase
$\exp(i \varepsilon \ldots)$, which is of the order of $1/\varepsilon$.
The smaller $\varepsilon$, the larger the smoothing of $P$ we
are allowed to do.
Suppose that the smoothing consists of substituting $P$ by 
a histogram of bin size $a \sim 1/\varepsilon$ for each coordinate
(see figure~\ref{fig45}).
Then, by choosing $a \ge 1/B$ the smoothed distribution 
becomes a constant, i.e., becomes separable (the marginals 
are constant). 
We conclude that for $\varepsilon \lesssim B$ the distribution $P$
is {\em effectively} separable and the random-dynamic result
is applicable.

Concerning the application of the transfer matrix method to 
more general maps with a finite symbolic dynamics, e.g., 
cat maps, it is likely that the scheme can also be adapted 
to this case. 
The general structure of the transfer matrix will be the same as 
that of the corresponding baker with the same number of symbols.
The novelty is that some zeros will appear along the staircases
whenever the associated symbolic sequence is prohibited (the 
information about allowed and prohibited sequences is contained
in the transition matrix \cite{adler98}).  
The correspondence between symbols and phase space points will 
be nontrivial, but it could be settled numerically, just by 
launching initial conditions and registering the sequence of
Markov regions that are visited.

The main obstacle we see at the moment is that in general
the symbolic 
sequences do not have the same weight.
It is possible that these weights can be embodied in the 
transfer matrix, or it may be a good approximation to assume
that for long sequences the weights are equal. 
This issue, which exceeds the original scope of the present 
paper, is currently being investigated \cite{pulpo_wip}.

\section{Crossover regime}
\label{sec5}
In the previous sections we provided analytical descriptions of 
the FA in two extreme cases:
the random dynamics limit and the random perturbation limit. 
The first model provides a satisfactory description for initial times 
in general and for 
all times -- before saturation -- in the large $\lambda$ limit (ideally $\lambda\to \infty$).
The second explains the appearance of the Lyapunov regime for 
long times and large enough perturbations. 

In many cases the time behavior of the FA is not pure, but shows 
a crossover from a random-dynamics decay (short times) to a
random-perturbation decay (long times). 
We know that for short times the influence of the dynamics is not 
significant on average, then the trajectories in the DR formula
(\ref{eq:ODRn}) behave like random walks with uncorrelated jumps. 
Therefore, for short times the decay of the average fidelity 
amplitude is given by \equa{eq:gamma2} --it is exact for $t=1$. 
In addition, for a general smooth perturbation with large amplitude, 
we have shown that there is a randomization of phases yielding an 
effectively random perturbation. 
As a consequence, the well known Lyapunov decay emerges 
in the large $\epsilon$ limit.
There 
remains a need for understanding the 
crossover (in the time domain) between random-dynamics and Lyapunov 
decays. 

In order to investigate the crossover we resorted to numerical 
simulations using the DR and the cat map (\ref{eq:pcat}) with a 
random 
perturbation. 
We divided the torus into $N_c\times N_c$ square cells, 
assigning a random constant value $v_i \in [-0.5,0.5]$ 
to each cell, with $i \in [1,N_c^2]$.
Next we 
computed $\overline{\ODRn}$ by summing over 
$n_s$ initial 
conditions and then averaged $|\overline{\ODRn}|$ over 
$N_{\rm rc}$ 
realizations of the perturbation.

\begin{figure}
\begin{center}
\includegraphics[width=0.9\linewidth]{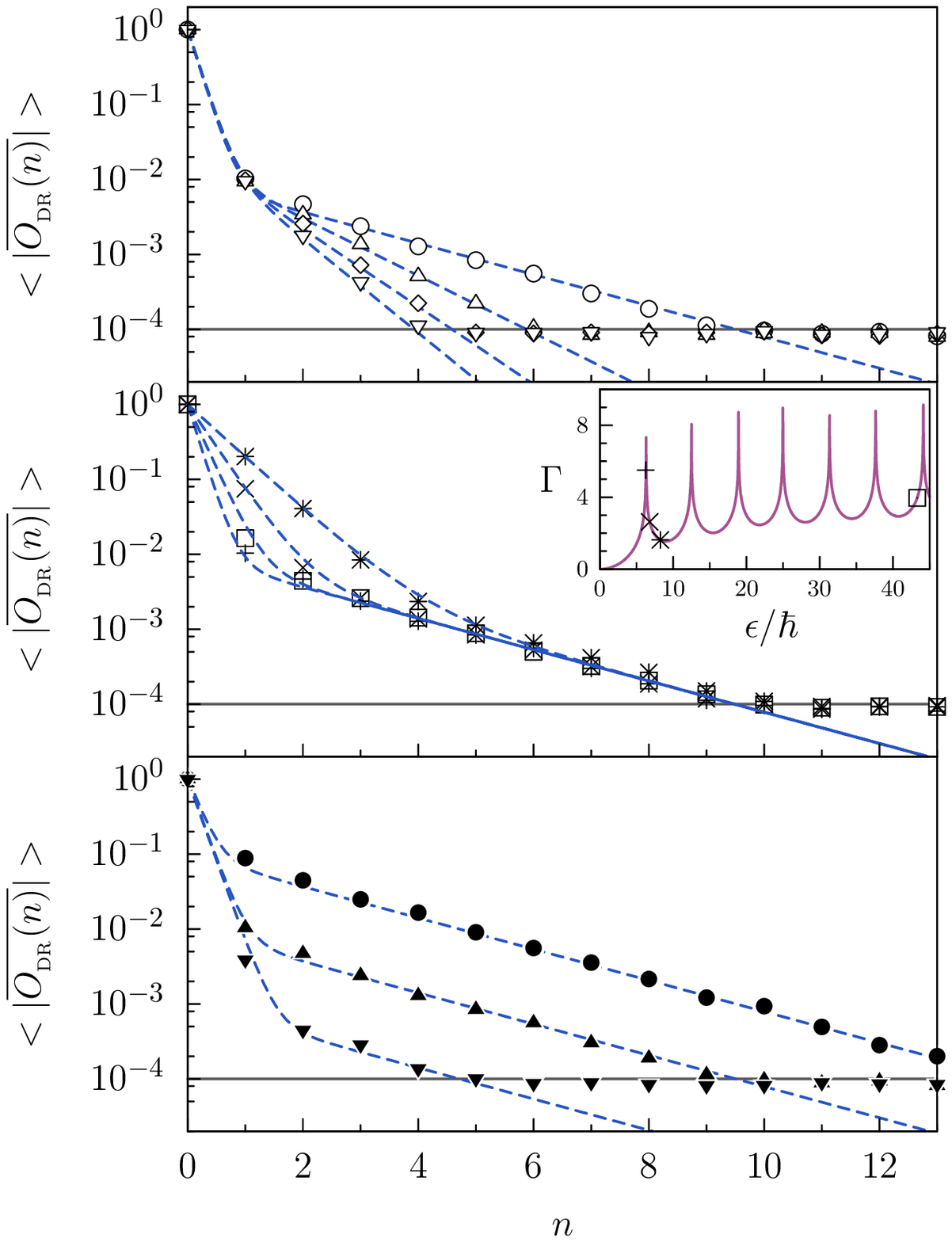} 
\caption{(color online) 
Top: $<|\overline{\ODRn}|>$ for $\epsilon/\hbar=6.2589$ 
($+$ symbol in the inset), 
and different values 
$a=b=1,\ 2,\ 3$ and $4$ in \equa{eq:pcat} 
giving 
($\circ$)       $\lambda=0.96$, 
($\triangle$)  $\lambda=1.76$, 
($\diamond$) $\lambda=2.389$ and 
($\triangledown$)       $\lambda=2.887$ 
respectively.
Middle:  
$<|\overline{\ODRn}|>$ for 
$a=b=1$, $\lambda=0.96$ and 
different perturbation strengths: 
($+$)       $\epsilon/\hbar=6.2589$; 
($\times$)  $\epsilon/\hbar=6.794$; 
($*$)       $\epsilon/\hbar=8.235$ ; 
($\square$) $\epsilon/\hbar=43.318$.
Inset: $\Gamma(t)$ computed from $N_{\rm rc}=100$ realizations 
of the perturbation with $N_c=100$. 
The points indicate the value of $\epsilon/\hbar$ for the curve 
with corresponding symbol on the bottom panel.
Bottom: $<|\overline{\ODRn}|>$ for 
$\epsilon/\hbar=6.2589$ ($+$ symbol in the inset), 
for three different values of $N_c$: 
($\bullet$)      $N_c=10$, 
($\blacktriangle$) $N_c=100$ and 
$(\blacktriangledown)$      $N_c=1000$
($a=b=1$, $\lambda=0.96$). 
The dashed blue line (in all panels) corresponds to the 
bi-exponential in \equa{eq:cross}.
The horizontal gray lines correspond to the expected saturation
values.
\label{fig:multi}}
\end{center}
\end{figure}

In figure~\ref{fig:multi} we show the results of our simulations 
for various values of $\epsilon$ and $\lambda$.
Using Eqs.~(\ref{eq:gamma1}) and (\ref{eq:gamma2}) we computed 
the decay rate $\Gamma$ for the initial times (see inset). 
We can already see on the top panel that $<|\overline{\ODRn}|>$ 
has three well defined regimes: two exponential decays followed 
by the expected saturation due to finite number of initial conditions. 
Then, up to the saturation point, the decay is well described by
\begin{equation}
\label{eq:cross}
 <|\overline{\ODRn}|> \sim \exp(-\Gamma n)+A \exp(-\lambda n/2) \, .
\end{equation}
We found empirically that, in our particular model, the prefactor 
$A$ is given by
\begin{equation}
A=\lambda/N_c \, .
\end{equation}
So, when $\lambda\to\infty$ the dynamics is random, the second term 
vanishes, and the decay rate is $\Gamma$ for all times before saturation, as predicted 
in Sec.~\ref{sec3}. 
When $\lambda$ is finite but the perturbation is completely random 
($N_c \to \infty$) the first term in Eq.~(\ref{eq:cross}) dominates
the initial decay.
The reason for this is that deterministic motion together with random 
perturbations looks, on short time scales, like random dynamics
(as far as the accumulation of phases is concerned).

In figure~\ref{fig:multi} (top and middle panels) the dependence of 
$A$ with $\lambda$ is clearly observed.
The dependence with $N_c$ is established in  
the bottom panel, where we plot $ <|\overline{\ODRn}|>$ for three 
different orders of magnitude of the randomness parameter $N_c$. 
We remark that, as expected, the crossover point depends both on 
the dynamics (through $\lambda$) and on the perturbation 
(through $N_c$).  
We can clearly see this in the middle panel of figure~\ref{fig:multi} where we 
 show $ <|\overline{\ODRn}|>$ for 
different values of the 
rescaled perturbation strength $\epsilon/\hbar$ 
[and for $\lambda=0.96$, corresponding to $a=b=1$ in \equa{eq:pcat}] and
in the bottom panel of the same figure where we plot  $ <|\overline{\ODRn}|>$ 
for one value of $\epsilon/\hbar$ and $N_c=10,\ 100$ and $1000$.
We can again see that \equa{eq:cross} describes quite well the decay 
up to the saturation value.

We thus verify that the crossover regime can be described by 
a natural bi-exponential law combining the random-dynamic and 
Lyapunov decays.

\section{Conclusions} 
\label{sec6}
We have analyzed the average fidelity amplitude, a quantity that 
characterizes the instability and irreversibility of perturbed 
quantum evolution.
We have taken as a starting point a well established  semiclassical
theory --the dephasing representation 
Our first key result is obtained in 
the limit of random dynamics. We  show that in that case the
decay of the average FA is exponential for all times -- up to the saturation --
and we give an analytic expression for the corresponding decay rate.
Maps on the torus with a very large Lyapunov exponent are expected to mimic 
random dynamics rather precisely. Thus the decay rate 
$\Gamma$ obtained in equation (\ref{eq:gamma2}) is to be observed for increasingly larger periods of time as 
$\lambda$ is increased. Numerical results shown in figures 1 and 3 confirm this assertion.

Our second key result is that using the DR and a novel approach in this context -- the transfer matrix method
-- we were able to predict that the asymptotic decay of the average fidelity amplitude should be 
perturbation independent and given by $\lambda/2$. We are even able to explain 
possible deviations from this behavior. 
We point out that the Lyapunov, as well as other regimes were obtained 
in \cite{vanicek2004} analyzing statistics of action differences in an expression which is obtained from 
the square of 
the average FA (equation (\ref{eq:afa})). However 
in that article the short time behavior due to random dynamics-like behavior is not observed. In addition, one major advantage of our approach is that we do not need to sum over pairs of trajectories -- a difficulty that arises
in the squaring needed to perform average fidelity studies.

 As consequence,  we have both analytical and numerical evidence to conclude that 
the average FA for an arbitrary  
chaotic map, should decay initially with a decay rate 
given by the random dynamics approximation, and for long times
there is the Lyapunov regime.
We cannot yet predict in a generic case the behavior or the time 
where the crossover takes place. Nevertheless in an attempt to understand this problem 
we have studied a
model of random perturbations which helped us in 
determining that for random perturbations and intermediate 
times the decay is also given by the Lyapunov exponent.
 
In \cite{nacho2011} it is shown that the decay of Loschmidt echo has 
three regimes as a function of the perturbation strength. 
The well known --quadratic-- Fermi golden rule regime is valid for 
small perturbations. 
Then there is a regime where the decay rate is given by $\Gamma$ 
(computed for random dynamics) and after that there
is a crossover to a perturbation independent regime given by 
the Lyapunov exponent $\lambda$.
We believe that this paper is an important step towards 
understanding this complex behavior of the LE.
We remark that the average FA is tightly related to the LE
but has some fundamental differences.
In fact, if we suppose that $n_s$ is the number of initial 
conditions then the average LE can be expressed as a sum of 
the square of the average of the FA (times $n_s$) minus 
a sum of ``nondiagonal'' terms consisting of 
products of $O(t)$ for different initial conditions. 
Here we have taken a significant step forward by understanding 
one of the parts of the LE: the average FA, which is in itself 
an important quantity in many experiments. 
In order to fully comprehend
the behavior of the LE further work is needed \cite{pulpo_wip}.
\section*{Acknowledgments}
We are grateful to F. Toscano for useful suggestions.
We received partial support from ANCyPT (grant PICT-2010-1556),
UBACyT (grant X237) (Argentina), CNPq (Brazil), and UNAM-PAPIIT 
(grant IN117310) (Mexico).

\begin{thebibliography}{49}%
\makeatletter
\providecommand \@ifxundefined [1]{%
 \@ifx{#1\undefined}
}%
\providecommand \@ifnum [1]{%
 \ifnum #1\expandafter \@firstoftwo
 \else \expandafter \@secondoftwo
 \fi
}%
\providecommand \@ifx [1]{%
 \ifx #1\expandafter \@firstoftwo
 \else \expandafter \@secondoftwo
 \fi
}%
\providecommand \natexlab [1]{#1}%
\providecommand \enquote  [1]{``#1''}%
\providecommand \bibnamefont  [1]{#1}%
\providecommand \bibfnamefont [1]{#1}%
\providecommand \citenamefont [1]{#1}%
\providecommand \href@noop [0]{\@secondoftwo}%
\providecommand \href [0]{\begingroup \@sanitize@url \@href}%
\providecommand \@href[1]{\@@startlink{#1}\@@href}%
\providecommand \@@href[1]{\endgroup#1\@@endlink}%
\providecommand \@sanitize@url [0]{\catcode `\\12\catcode `\$12\catcode
  `\&12\catcode `\#12\catcode `\^12\catcode `\_12\catcode `\%12\relax}%
\providecommand \@@startlink[1]{}%
\providecommand \@@endlink[0]{}%
\providecommand \url  [0]{\begingroup\@sanitize@url \@url }%
\providecommand \@url [1]{\endgroup\@href {#1}{\urlprefix }}%
\providecommand \urlprefix  [0]{URL }%
\providecommand \Eprint [0]{\href }%
\providecommand \doibase [0]{http://dx.doi.org/}%
\providecommand \selectlanguage [0]{\@gobble}%
\providecommand \bibinfo  [0]{\@secondoftwo}%
\providecommand \bibfield  [0]{\@secondoftwo}%
\providecommand \translation [1]{[#1]}%
\providecommand \BibitemOpen [0]{}%
\providecommand \bibitemStop [0]{}%
\providecommand \bibitemNoStop [0]{.\EOS\space}%
\providecommand \EOS [0]{\spacefactor3000\relax}%
\providecommand \BibitemShut  [1]{\csname bibitem#1\endcsname}%
\let\auto@bib@innerbib\@empty
\bibitem [{\citenamefont {Peres}(1984)}]{Peres1984}%
  \BibitemOpen
  \bibfield  {author} {\bibinfo {author} {\bibfnamefont {A.}~\bibnamefont
  {Peres}},\ }\href {\doibase 10.1103/PhysRevA.30.1610} {\bibfield  {journal}
  {\bibinfo  {journal} {Phys. Rev. A}\ }\textbf {\bibinfo {volume} {30}},\
  \bibinfo {pages} {1610} (\bibinfo {year} {1984})}\BibitemShut {NoStop}%
\bibitem [{\citenamefont {Jalabert}\ and\ \citenamefont
  {Pastawski}(2001)}]{Jalabert2001}%
  \BibitemOpen
  \bibfield  {author} {\bibinfo {author} {\bibfnamefont {R.~A.}\ \bibnamefont
  {Jalabert}}\ and\ \bibinfo {author} {\bibfnamefont {H.~M.}\ \bibnamefont
  {Pastawski}},\ }\href {\doibase 10.1103/PhysRevLett.86.2490} {\bibfield
  {journal} {\bibinfo  {journal} {Phys. Rev. Lett.}\ }\textbf {\bibinfo
  {volume} {86}},\ \bibinfo {pages} {2490} (\bibinfo {year}
  {2001})}\BibitemShut {NoStop}%
\bibitem [{\citenamefont {\mbox{Ph.} Jacquod}\ \emph
  {et~al.}(2001)\citenamefont {\mbox{Ph.} Jacquod}, \citenamefont
  {Silvestrov},\ and\ \citenamefont {Beenakker}}]{Jacquod2001}%
  \BibitemOpen
  \bibfield  {author} {\bibinfo {author} {\bibnamefont {\mbox{Ph.} Jacquod}},
  \bibinfo {author} {\bibfnamefont {P.~G.}\ \bibnamefont {Silvestrov}}, \ and\
  \bibinfo {author} {\bibfnamefont {C.~W.~J.}\ \bibnamefont {Beenakker}},\
  }\href {\doibase 10.1103/PhysRevE.64.055203} {\bibfield  {journal} {\bibinfo
  {journal} {Phys. Rev. E}\ }\textbf {\bibinfo {volume} {64}},\ \bibinfo
  {pages} {055203} (\bibinfo {year} {2001})}\BibitemShut {NoStop}%
\bibitem [{\citenamefont {Gorin}\ \emph
  {et~al.}(2006{\natexlab{a}})\citenamefont {Gorin}, \citenamefont {Prosen},
  \citenamefont {Seligman},\ and\ \citenamefont {{\v Z}nidari{\v
  c}}}]{Gorin2006}%
  \BibitemOpen
  \bibfield  {author} {\bibinfo {author} {\bibfnamefont {T.}~\bibnamefont
  {Gorin}}, \bibinfo {author} {\bibfnamefont {T.}~\bibnamefont {Prosen}},
  \bibinfo {author} {\bibfnamefont {T.}~\bibnamefont {Seligman}}, \ and\
  \bibinfo {author} {\bibfnamefont {M.}~\bibnamefont {{\v Z}nidari{\v c}}},\
  }\href@noop {} {\bibfield  {journal} {\bibinfo  {journal} {Phys. Rep.}\
  }\textbf {\bibinfo {volume} {435}},\ \bibinfo {pages} {33} (\bibinfo {year}
  {2006}{\natexlab{a}})}\BibitemShut {NoStop}%
\bibitem [{\citenamefont {\mbox{Ph.} Jacquod}\ and\ \citenamefont
  {Petitjean}(2009)}]{Jacquod2009}%
  \BibitemOpen
  \bibfield  {author} {\bibinfo {author} {\bibnamefont {\mbox{Ph.} Jacquod}}\
  and\ \bibinfo {author} {\bibfnamefont {C.}~\bibnamefont {Petitjean}},\
  }\href@noop {} {\bibfield  {journal} {\bibinfo  {journal} {Adv. Phys.}\
  }\textbf {\bibinfo {volume} {58}},\ \bibinfo {pages} {67} (\bibinfo {year}
  {2009})}\BibitemShut {NoStop}%
\bibitem [{\citenamefont {Prosen}\ and\ \citenamefont {{\v Z}nidari{\v
  c}}(2002)}]{Prosen2002}%
  \BibitemOpen
  \bibfield  {author} {\bibinfo {author} {\bibfnamefont {T.}~\bibnamefont
  {Prosen}}\ and\ \bibinfo {author} {\bibfnamefont {M.}~\bibnamefont {{\v
  Z}nidari{\v c}}},\ }\href@noop {} {\bibfield  {journal} {\bibinfo  {journal}
  {J. Phys. A: Math. Gen.}\ }\textbf {\bibinfo {volume} {35}},\ \bibinfo
  {pages} {1455} (\bibinfo {year} {2002})}\BibitemShut {NoStop}%
\bibitem [{\citenamefont {Prosen}(2002)}]{Prosen2002_2}%
  \BibitemOpen
  \bibfield  {author} {\bibinfo {author} {\bibfnamefont {T.}~\bibnamefont
  {Prosen}},\ }\href@noop {} {\bibfield  {journal} {\bibinfo  {journal} {Phys.
  Rev. E}\ }\textbf {\bibinfo {volume} {65}},\ \bibinfo {pages} {036208}
  (\bibinfo {year} {2002})}\BibitemShut {NoStop}%
\bibitem [{\citenamefont {Wang}\ \emph {et~al.}(2004)\citenamefont {Wang},
  \citenamefont {Casati},\ and\ \citenamefont {Li}}]{Wang2004}%
  \BibitemOpen
  \bibfield  {author} {\bibinfo {author} {\bibfnamefont {W.}~\bibnamefont
  {Wang}}, \bibinfo {author} {\bibfnamefont {G.}~\bibnamefont {Casati}}, \ and\
  \bibinfo {author} {\bibfnamefont {B.}~\bibnamefont {Li}},\ }\href {\doibase
  10.1103/PhysRevE.69.025201} {\bibfield  {journal} {\bibinfo  {journal} {Phys.
  Rev. E}\ }\textbf {\bibinfo {volume} {69}},\ \bibinfo {pages} {025201}
  (\bibinfo {year} {2004})}\BibitemShut {NoStop}%
\bibitem [{\citenamefont {Andersen}\ \emph {et~al.}(2006)\citenamefont
  {Andersen}, \citenamefont {Kaplan}, \citenamefont {Gr{\"u}nzweig},\ and\
  \citenamefont {Davidson}}]{Andersen2006}%
  \BibitemOpen
  \bibfield  {author} {\bibinfo {author} {\bibfnamefont {M.}~\bibnamefont
  {Andersen}}, \bibinfo {author} {\bibfnamefont {A.}~\bibnamefont {Kaplan}},
  \bibinfo {author} {\bibfnamefont {T.}~\bibnamefont {Gr{\"u}nzweig}}, \ and\
  \bibinfo {author} {\bibfnamefont {N.}~\bibnamefont {Davidson}},\ }\href@noop
  {} {\bibfield  {journal} {\bibinfo  {journal} {Phys. Rev. Lett.}\ }\textbf
  {\bibinfo {volume} {97}},\ \bibinfo {pages} {104102} (\bibinfo {year}
  {2006})}\BibitemShut {NoStop}%
\bibitem [{\citenamefont {Pozzo}\ and\ \citenamefont
  {Dom{\'\i}nguez}(2007)}]{Pozzo2007}%
  \BibitemOpen
  \bibfield  {author} {\bibinfo {author} {\bibfnamefont {E.~N.}\ \bibnamefont
  {Pozzo}}\ and\ \bibinfo {author} {\bibfnamefont {D.}~\bibnamefont
  {Dom{\'\i}nguez}},\ }\href@noop {} {\bibfield  {journal} {\bibinfo  {journal}
  {Phys. Rev. Lett.}\ }\textbf {\bibinfo {volume} {98}},\ \bibinfo {pages}
  {057006} (\bibinfo {year} {2007})}\BibitemShut {NoStop}%
\bibitem [{\citenamefont {Ares}\ and\ \citenamefont
  {Wisniacki}(2009)}]{Natalia2009}%
  \BibitemOpen
  \bibfield  {author} {\bibinfo {author} {\bibfnamefont {N.}~\bibnamefont
  {Ares}}\ and\ \bibinfo {author} {\bibfnamefont {D.~A.}\ \bibnamefont
  {Wisniacki}},\ }\href@noop {} {\bibfield  {journal} {\bibinfo  {journal}
  {Phys. Rev. E}\ }\textbf {\bibinfo {volume} {80}},\ \bibinfo {pages} {046216}
  (\bibinfo {year} {2009})}\BibitemShut {NoStop}%
\bibitem [{\citenamefont {Casabone}\ \emph {et~al.}(2010)\citenamefont
  {Casabone}, \citenamefont {Garc{\'\i}a-Mata},\ and\ \citenamefont
  {Wisniacki}}]{Casabone2010}%
  \BibitemOpen
  \bibfield  {author} {\bibinfo {author} {\bibfnamefont {B.}~\bibnamefont
  {Casabone}}, \bibinfo {author} {\bibfnamefont {I.}~\bibnamefont
  {Garc{\'\i}a-Mata}}, \ and\ \bibinfo {author} {\bibfnamefont
  {D.}~\bibnamefont {Wisniacki}},\ }\href {\doibase 10.1209/0295-5075/89/50009}
  {\bibfield  {journal} {\bibinfo  {journal} {Europhys. Lett.}\ }\textbf
  {\bibinfo {volume} {89}},\ \bibinfo {pages} {50009} (\bibinfo {year}
  {2010})}\BibitemShut {NoStop}%
\bibitem [{\citenamefont {Lobkis}\ and\ \citenamefont
  {Weaver}(2003)}]{Lobkis2003}%
  \BibitemOpen
  \bibfield  {author} {\bibinfo {author} {\bibfnamefont {O.~I.}\ \bibnamefont
  {Lobkis}}\ and\ \bibinfo {author} {\bibfnamefont {R.~L.}\ \bibnamefont
  {Weaver}},\ }\href {\doibase 10.1103/PhysRevLett.90.254302} {\bibfield
  {journal} {\bibinfo  {journal} {Phys. Rev. Lett.}\ }\textbf {\bibinfo
  {volume} {90}},\ \bibinfo {pages} {254302} (\bibinfo {year}
  {2003})}\BibitemShut {NoStop}%
\bibitem [{\citenamefont {Sch\"afer}\ \emph
  {et~al.}(2005{\natexlab{a}})\citenamefont {Sch\"afer}, \citenamefont
  {St\"ockmann}, \citenamefont {Gorin},\ and\ \citenamefont
  {Seligman}}]{Schafer2005}%
  \BibitemOpen
  \bibfield  {author} {\bibinfo {author} {\bibfnamefont {R.}~\bibnamefont
  {Sch\"afer}}, \bibinfo {author} {\bibfnamefont {H.-J.}\ \bibnamefont
  {St\"ockmann}}, \bibinfo {author} {\bibfnamefont {T.}~\bibnamefont {Gorin}},
  \ and\ \bibinfo {author} {\bibfnamefont {T.~H.}\ \bibnamefont {Seligman}},\
  }\href@noop {} {\bibfield  {journal} {\bibinfo  {journal} {Phys. Rev. Lett.}\
  }\textbf {\bibinfo {volume} {95}},\ \bibinfo {pages} {184102} (\bibinfo
  {year} {2005}{\natexlab{a}})}\BibitemShut {NoStop}%
\bibitem [{\citenamefont {Sch\"afer}\ \emph
  {et~al.}(2005{\natexlab{b}})\citenamefont {Sch\"afer}, \citenamefont {Gorin},
  \citenamefont {Seligman},\ and\ \citenamefont {St\"ockmann}}]{Schafer2005_2}%
  \BibitemOpen
  \bibfield  {author} {\bibinfo {author} {\bibfnamefont {R.}~\bibnamefont
  {Sch\"afer}}, \bibinfo {author} {\bibfnamefont {T.}~\bibnamefont {Gorin}},
  \bibinfo {author} {\bibfnamefont {T.~H.}\ \bibnamefont {Seligman}}, \ and\
  \bibinfo {author} {\bibfnamefont {H.-J.}\ \bibnamefont {St\"ockmann}},\
  }\href@noop {} {\bibfield  {journal} {\bibinfo  {journal} {New J. Phys.}\
  }\textbf {\bibinfo {volume} {7}},\ \bibinfo {pages} {152} (\bibinfo {year}
  {2005}{\natexlab{b}})}\BibitemShut {NoStop}%
\bibitem [{\citenamefont {Gorin}\ \emph
  {et~al.}(2006{\natexlab{b}})\citenamefont {Gorin}, \citenamefont {Seligman},\
  and\ \citenamefont {Weaver}}]{Gorin2006_2}%
  \BibitemOpen
  \bibfield  {author} {\bibinfo {author} {\bibfnamefont {T.}~\bibnamefont
  {Gorin}}, \bibinfo {author} {\bibfnamefont {T.~H.}\ \bibnamefont {Seligman}},
  \ and\ \bibinfo {author} {\bibfnamefont {R.~L.}\ \bibnamefont {Weaver}},\
  }\href {\doibase 10.1103/PhysRevE.73.015202} {\bibfield  {journal} {\bibinfo
  {journal} {Phys. Rev. E}\ }\textbf {\bibinfo {volume} {73}},\ \bibinfo
  {pages} {015202} (\bibinfo {year} {2006}{\natexlab{b}})}\BibitemShut
  {NoStop}%
\bibitem [{\citenamefont {Lobkis}\ and\ \citenamefont
  {Weaver}(2008)}]{Lobkis2008}%
  \BibitemOpen
  \bibfield  {author} {\bibinfo {author} {\bibfnamefont {O.~I.}\ \bibnamefont
  {Lobkis}}\ and\ \bibinfo {author} {\bibfnamefont {R.~L.}\ \bibnamefont
  {Weaver}},\ }\href {\doibase 10.1103/PhysRevE.78.066212} {\bibfield
  {journal} {\bibinfo  {journal} {Phys. Rev. E}\ }\textbf {\bibinfo {volume}
  {78}},\ \bibinfo {pages} {066212} (\bibinfo {year} {2008})}\BibitemShut
  {NoStop}%
\bibitem [{\citenamefont {K\"ober}\ \emph {et~al.}(2010)\citenamefont
  {K\"ober}, \citenamefont {Kuhl}, \citenamefont {St\"ockmann}, \citenamefont
  {Gorin}, \citenamefont {Savin},\ and\ \citenamefont {Seligman}}]{Kober2010}%
  \BibitemOpen
  \bibfield  {author} {\bibinfo {author} {\bibfnamefont {B.}~\bibnamefont
  {K\"ober}}, \bibinfo {author} {\bibfnamefont {U.}~\bibnamefont {Kuhl}},
  \bibinfo {author} {\bibfnamefont {H.-J.}\ \bibnamefont {St\"ockmann}},
  \bibinfo {author} {\bibfnamefont {T.}~\bibnamefont {Gorin}}, \bibinfo
  {author} {\bibfnamefont {D.~V.}\ \bibnamefont {Savin}}, \ and\ \bibinfo
  {author} {\bibfnamefont {T.~H.}\ \bibnamefont {Seligman}},\ }\href {\doibase
  10.1103/PhysRevE.82.036207} {\bibfield  {journal} {\bibinfo  {journal} {Phys.
  Rev. E}\ }\textbf {\bibinfo {volume} {82}},\ \bibinfo {pages} {036207}
  (\bibinfo {year} {2010})}\BibitemShut {NoStop}%
\bibitem [{\citenamefont {Van\'i\ifmmode~\check{c}\else
  \v{c}\fi{}ek}(2004)}]{vanicek2004}%
  \BibitemOpen
  \bibfield  {author} {\bibinfo {author} {\bibfnamefont {J.}~\bibnamefont
  {Van\'i\ifmmode~\check{c}\else \v{c}\fi{}ek}},\ }\href@noop {} {\bibfield
  {journal} {\bibinfo  {journal} {Phys. Rev. E}\ }\textbf {\bibinfo {volume}
  {70}},\ \bibinfo {pages} {055201} (\bibinfo {year} {2004})}\BibitemShut
  {NoStop}%
\bibitem [{\citenamefont {Van\'i\ifmmode~\check{c}\else \v{c}\fi{}ek}\ and\
  \citenamefont {Heller}(2003)}]{vanicek2003}%
  \BibitemOpen
  \bibfield  {author} {\bibinfo {author} {\bibfnamefont {J.}~\bibnamefont
  {Van\'i\ifmmode~\check{c}\else \v{c}\fi{}ek}}\ and\ \bibinfo {author}
  {\bibfnamefont {E.~J.}\ \bibnamefont {Heller}},\ }\href@noop {} {\bibfield
  {journal} {\bibinfo  {journal} {Phys. Rev. E}\ }\textbf {\bibinfo {volume}
  {68}},\ \bibinfo {pages} {056208} (\bibinfo {year} {2003})}\BibitemShut
  {NoStop}%
\bibitem [{\citenamefont {Van{\'\i}{\v c}ek}(2006)}]{vanicek2006}%
  \BibitemOpen
  \bibfield  {author} {\bibinfo {author} {\bibfnamefont {J.}~\bibnamefont
  {Van{\'\i}{\v c}ek}},\ }\href@noop {} {\bibfield  {journal} {\bibinfo
  {journal} {Phys. Rev. E}\ }\textbf {\bibinfo {volume} {73}},\ \bibinfo
  {pages} {046204} (\bibinfo {year} {2006})}\BibitemShut {NoStop}%
\bibitem [{\citenamefont {Garc\'ia-Mata}\ and\ \citenamefont
  {Wisniacki}(2011)}]{nacho2011}%
  \BibitemOpen
  \bibfield  {author} {\bibinfo {author} {\bibfnamefont {I.}~\bibnamefont
  {Garc\'ia-Mata}}\ and\ \bibinfo {author} {\bibfnamefont {D.~A.}\ \bibnamefont
  {Wisniacki}},\ }\href@noop {} {\bibfield  {journal} {\bibinfo  {journal} {J.
  Phys. A: Math. Theor.}\ }\textbf {\bibinfo {volume} {44}},\ \bibinfo {pages}
  {315101} (\bibinfo {year} {2011})}\BibitemShut {NoStop}%
\bibitem [{\citenamefont {Miller}(1970)}]{Miller1970}%
  \BibitemOpen
  \bibfield  {author} {\bibinfo {author} {\bibfnamefont {W.~H.}\ \bibnamefont
  {Miller}},\ }\href@noop {} {\bibfield  {journal} {\bibinfo  {journal} {J.
  Chem. Phys.}\ }\textbf {\bibinfo {volume} {53}},\ \bibinfo {pages} {3578}
  (\bibinfo {year} {1970})}\BibitemShut {NoStop}%
\bibitem [{\citenamefont {Miller}(2001)}]{Miller2001}%
  \BibitemOpen
  \bibfield  {author} {\bibinfo {author} {\bibfnamefont {W.~H.}\ \bibnamefont
  {Miller}},\ }\href@noop {} {\bibfield  {journal} {\bibinfo  {journal} {J.
  Phys. Chem}\ }\textbf {\bibinfo {volume} {105}},\ \bibinfo {pages} {2942}
  (\bibinfo {year} {2001})}\BibitemShut {NoStop}%
\bibitem [{\citenamefont {Van{\'\i}{\v c}ek}(2004)}]{vanicek2004arxiv}%
  \BibitemOpen
  \bibfield  {author} {\bibinfo {author} {\bibfnamefont {J.}~\bibnamefont
  {Van{\'\i}{\v c}ek}},\ }\href@noop {} {} (\bibinfo {year} {2004}),\ \Eprint
  {http://arxiv.org/abs/arXiv:quant-ph/0410205} {arXiv:arXiv:quant-ph/0410205}
  \BibitemShut {NoStop}%
\bibitem [{\citenamefont {Wang}\ \emph {et~al.}(2005)\citenamefont {Wang},
  \citenamefont {Casati}, \citenamefont {Li},\ and\ \citenamefont
  {Prosen}}]{wang2005}%
  \BibitemOpen
  \bibfield  {author} {\bibinfo {author} {\bibfnamefont {W.}~\bibnamefont
  {Wang}}, \bibinfo {author} {\bibfnamefont {G.}~\bibnamefont {Casati}},
  \bibinfo {author} {\bibfnamefont {B.}~\bibnamefont {Li}}, \ and\ \bibinfo
  {author} {\bibfnamefont {T.}~\bibnamefont {Prosen}},\ }\href@noop {}
  {\bibfield  {journal} {\bibinfo  {journal} {Phys. Rev. E}\ }\textbf {\bibinfo
  {volume} {71}},\ \bibinfo {pages} {037202} (\bibinfo {year}
  {2005})}\BibitemShut {NoStop}%
\bibitem [{\citenamefont {Li}\ \emph {et~al.}(2009)\citenamefont {Li},
  \citenamefont {Mollica},\ and\ \citenamefont {Van{\'\i}{\v c}ek}}]{Li2009}%
  \BibitemOpen
  \bibfield  {author} {\bibinfo {author} {\bibfnamefont {B.}~\bibnamefont
  {Li}}, \bibinfo {author} {\bibfnamefont {C.}~\bibnamefont {Mollica}}, \ and\
  \bibinfo {author} {\bibfnamefont {J.}~\bibnamefont {Van{\'\i}{\v c}ek}},\
  }\href@noop {} {\bibfield  {journal} {\bibinfo  {journal} {J. Chem. Phys.}\
  }\textbf {\bibinfo {volume} {131}},\ \bibinfo {pages} {041101} (\bibinfo
  {year} {2009})}\BibitemShut {NoStop}%
\bibitem [{\citenamefont {Zimmermann}\ and\ \citenamefont {Van{\'\i}{\v
  c}ek}(2010)}]{Zimm2010}%
  \BibitemOpen
  \bibfield  {author} {\bibinfo {author} {\bibfnamefont {T.}~\bibnamefont
  {Zimmermann}}\ and\ \bibinfo {author} {\bibfnamefont {J.}~\bibnamefont
  {Van{\'\i}{\v c}ek}},\ }\href@noop {} {\bibfield  {journal} {\bibinfo
  {journal} {J. Chem. Phys.}\ }\textbf {\bibinfo {volume} {132}},\ \bibinfo
  {pages} {241101} (\bibinfo {year} {2010})}\BibitemShut {NoStop}%
\bibitem [{\citenamefont {Zimmermann}\ \emph {et~al.}(2010)\citenamefont
  {Zimmermann}, \citenamefont {Ruppen}, \citenamefont {Li},\ and\ \citenamefont
  {Van{\'\i}{\v c}ek}}]{Zimm2010_2}%
  \BibitemOpen
  \bibfield  {author} {\bibinfo {author} {\bibfnamefont {T.}~\bibnamefont
  {Zimmermann}}, \bibinfo {author} {\bibfnamefont {J.}~\bibnamefont {Ruppen}},
  \bibinfo {author} {\bibfnamefont {B.}~\bibnamefont {Li}}, \ and\ \bibinfo
  {author} {\bibfnamefont {J.}~\bibnamefont {Van{\'\i}{\v c}ek}},\ }\href@noop
  {} {\bibfield  {journal} {\bibinfo  {journal} {Int. J. Quant. Chem.}\
  }\textbf {\bibinfo {volume} {110}},\ \bibinfo {pages} {2426} (\bibinfo {year}
  {2010})}\BibitemShut {NoStop}%
\bibitem [{\citenamefont {Wehrle}\ \emph {et~al.}(2011)\citenamefont {Wehrle},
  \citenamefont {{\v S}ulc},\ and\ \citenamefont {Van{\'\i}{\v
  c}ek}}]{Wehrle2011}%
  \BibitemOpen
  \bibfield  {author} {\bibinfo {author} {\bibfnamefont {M.}~\bibnamefont
  {Wehrle}}, \bibinfo {author} {\bibfnamefont {M.}~\bibnamefont {{\v S}ulc}}, \
  and\ \bibinfo {author} {\bibfnamefont {J.}~\bibnamefont {Van{\'\i}{\v
  c}ek}},\ }\href@noop {} {\bibfield  {journal} {\bibinfo  {journal} {CHIMIA
  International Journal for Chemistry}\ }\textbf {\bibinfo {volume} {65}},\
  \bibinfo {pages} {334} (\bibinfo {year} {2011})}\BibitemShut {NoStop}%
\bibitem [{\citenamefont {Shi}\ and\ \citenamefont {Geva}(2005)}]{Shi2005}%
  \BibitemOpen
  \bibfield  {author} {\bibinfo {author} {\bibfnamefont {Q.}~\bibnamefont
  {Shi}}\ and\ \bibinfo {author} {\bibfnamefont {E.}~\bibnamefont {Geva}},\
  }\href@noop {} {\bibfield  {journal} {\bibinfo  {journal} {J. Chem. Phys}\
  }\textbf {\bibinfo {volume} {122}},\ \bibinfo {pages} {064506} (\bibinfo
  {year} {2005})}\BibitemShut {NoStop}%
\bibitem [{\citenamefont {Haake}(2010)}]{haake}%
  \BibitemOpen
  \bibfield  {author} {\bibinfo {author} {\bibfnamefont {F.}~\bibnamefont
  {Haake}},\ }\href@noop {} {\emph {\bibinfo {title} {Quantum Signatures of
  Chaos}}}\ (\bibinfo  {publisher} {{Springer-Verlag, Berlin}},\ \bibinfo
  {year} {2010})\BibitemShut {NoStop}%
\bibitem [{\citenamefont {de~Almeida}(1988)}]{ozorio}%
  \BibitemOpen
  \bibfield  {author} {\bibinfo {author} {\bibfnamefont {A.~M.~O.}\
  \bibnamefont {de~Almeida}},\ }\href@noop {} {}\ (\bibinfo  {publisher}
  {Cambridge University Press, Cambridge, UK},\ \bibinfo {year}
  {1988})\BibitemShut {NoStop}%
\bibitem [{\citenamefont {de~Matos}\ and\ \citenamefont
  {de~Almeida}(1995)}]{dematos}%
  \BibitemOpen
  \bibfield  {author} {\bibinfo {author} {\bibfnamefont {M.~B.}\ \bibnamefont
  {de~Matos}}\ and\ \bibinfo {author} {\bibfnamefont {A.~M.~O.}\ \bibnamefont
  {de~Almeida}},\ }\href@noop {} {\bibfield  {journal} {\bibinfo  {journal}
  {Ann. Phys.}\ }\textbf {\bibinfo {volume} {237}},\ \bibinfo {pages} {46}
  (\bibinfo {year} {1995})}\BibitemShut {NoStop}%
\bibitem [{\citenamefont {Guti\'errez}\ and\ \citenamefont
  {Goussev}(2009)}]{gutierrez2009}%
  \BibitemOpen
  \bibfield  {author} {\bibinfo {author} {\bibfnamefont {M.}~\bibnamefont
  {Guti\'errez}}\ and\ \bibinfo {author} {\bibfnamefont {A.}~\bibnamefont
  {Goussev}},\ }\href {\doibase 10.1103/PhysRevE.79.046211} {\bibfield
  {journal} {\bibinfo  {journal} {Phys. Rev. E}\ }\textbf {\bibinfo {volume}
  {79}},\ \bibinfo {pages} {046211} (\bibinfo {year} {2009})}\BibitemShut
  {NoStop}%
\bibitem [{\citenamefont {Goussev}\ \emph {et~al.}(2008)\citenamefont
  {Goussev}, \citenamefont {Waltner}, \citenamefont {Richter},\ and\
  \citenamefont {Jalabert}}]{Goussev2008}%
  \BibitemOpen
  \bibfield  {author} {\bibinfo {author} {\bibfnamefont {A.}~\bibnamefont
  {Goussev}}, \bibinfo {author} {\bibfnamefont {D.}~\bibnamefont {Waltner}},
  \bibinfo {author} {\bibfnamefont {K.}~\bibnamefont {Richter}}, \ and\
  \bibinfo {author} {\bibfnamefont {R.~A.}\ \bibnamefont {Jalabert}},\
  }\href@noop {} {\bibfield  {journal} {\bibinfo  {journal} {New J. Phys.}\
  }\textbf {\bibinfo {volume} {10}},\ \bibinfo {pages} {093010} (\bibinfo
  {year} {2008})}\BibitemShut {NoStop}%
\bibitem [{\citenamefont {Dobson}(1969)}]{dobson69}%
  \BibitemOpen
  \bibfield  {author} {\bibinfo {author} {\bibfnamefont {J.~F.}\ \bibnamefont
  {Dobson}},\ }\href@noop {} {\bibfield  {journal} {\bibinfo  {journal} {J.
  Math. Phys.}\ }\textbf {\bibinfo {volume} {10}},\ \bibinfo {pages} {40}
  (\bibinfo {year} {1969})}\BibitemShut {NoStop}%
\bibitem [{\citenamefont {Borzi}\ \emph {et~al.}(1987)\citenamefont {Borzi},
  \citenamefont {Ord},\ and\ \citenamefont {Percus}}]{borzi87}%
  \BibitemOpen
  \bibfield  {author} {\bibinfo {author} {\bibfnamefont {C.}~\bibnamefont
  {Borzi}}, \bibinfo {author} {\bibfnamefont {G.}~\bibnamefont {Ord}}, \ and\
  \bibinfo {author} {\bibfnamefont {J.~K.}\ \bibnamefont {Percus}},\
  }\href@noop {} {\bibfield  {journal} {\bibinfo  {journal} {J. Stat. Phys.}\
  }\textbf {\bibinfo {volume} {46}},\ \bibinfo {pages} {51} (\bibinfo {year}
  {1987})}\BibitemShut {NoStop}%
\bibitem [{\citenamefont {Reichl}(1998)}]{reichl}%
  \BibitemOpen
  \bibfield  {author} {\bibinfo {author} {\bibfnamefont {L.~E.}\ \bibnamefont
  {Reichl}},\ }\href@noop {} {\emph {\bibinfo {title} {A Modern Course in
  Statistical Physics}}}\ (\bibinfo  {publisher} {{John Wiley \& Sons, Inc.,
  New York}},\ \bibinfo {year} {1998})\BibitemShut {NoStop}%
\bibitem [{\citenamefont {Gutzwiller}(1990)}]{gutzwiller}%
  \BibitemOpen
  \bibfield  {author} {\bibinfo {author} {\bibfnamefont {M.~C.}\ \bibnamefont
  {Gutzwiller}},\ }\href@noop {} {\emph {\bibinfo {title} {Chaos in Classical
  ad Quantum Mechanics}}}\ (\bibinfo  {publisher} {{Springer, New York}},\
  \bibinfo {year} {1990})\BibitemShut {NoStop}%
\bibitem [{\citenamefont {Kaplan}\ and\ \citenamefont
  {Heller}(1996)}]{kaplan96}%
  \BibitemOpen
  \bibfield  {author} {\bibinfo {author} {\bibfnamefont {L.}~\bibnamefont
  {Kaplan}}\ and\ \bibinfo {author} {\bibfnamefont {E.~J.}\ \bibnamefont
  {Heller}},\ }\href {\doibase 10.1103/PhysRevLett.76.1453} {\bibfield
  {journal} {\bibinfo  {journal} {Phys. Rev. Lett.}\ }\textbf {\bibinfo
  {volume} {76}},\ \bibinfo {pages} {1453} (\bibinfo {year}
  {1996})}\BibitemShut {NoStop}%
\bibitem [{\citenamefont {Smilansky}\ and\ \citenamefont
  {Verdene}(2003)}]{smilansky03}%
  \BibitemOpen
  \bibfield  {author} {\bibinfo {author} {\bibfnamefont {U.}~\bibnamefont
  {Smilansky}}\ and\ \bibinfo {author} {\bibfnamefont {B.}~\bibnamefont
  {Verdene}},\ }\href@noop {} {\bibfield  {journal} {\bibinfo  {journal} {J.
  Phys. A: Math. Gen.}\ }\textbf {\bibinfo {volume} {36}},\ \bibinfo {pages}
  {3525} (\bibinfo {year} {2003})}\BibitemShut {NoStop}%
\bibitem [{\citenamefont {Carlo}\ \emph {et~al.}(2010)\citenamefont {Carlo},
  \citenamefont {Vallejos},\ and\ \citenamefont {Abreu}}]{carlo10}%
  \BibitemOpen
  \bibfield  {author} {\bibinfo {author} {\bibfnamefont {G.~G.}\ \bibnamefont
  {Carlo}}, \bibinfo {author} {\bibfnamefont {R.~O.}\ \bibnamefont {Vallejos}},
  \ and\ \bibinfo {author} {\bibfnamefont {R.~F.}\ \bibnamefont {Abreu}},\
  }\href {\doibase 10.1103/PhysRevE.82.046220} {\bibfield  {journal} {\bibinfo
  {journal} {Phys. Rev. E}\ }\textbf {\bibinfo {volume} {82}},\ \bibinfo
  {pages} {046220} (\bibinfo {year} {2010})}\BibitemShut {NoStop}%
\bibitem [{\citenamefont {Arnold}\ and\ \citenamefont {Avez}(1989)}]{arnold}%
  \BibitemOpen
  \bibfield  {author} {\bibinfo {author} {\bibfnamefont {V.~I.}\ \bibnamefont
  {Arnold}}\ and\ \bibinfo {author} {\bibfnamefont {A.}~\bibnamefont {Avez}},\
  }\href@noop {} {\emph {\bibinfo {title} {Ergodic Problems in Classical
  Mechanics}}}\ (\bibinfo  {publisher} {{Addison-Wesley, Reading}},\ \bibinfo
  {year} {1989})\BibitemShut {NoStop}%
\bibitem [{\citenamefont {Adler}(1998)}]{adler98}%
  \BibitemOpen
  \bibfield  {author} {\bibinfo {author} {\bibfnamefont {R.~L.}\ \bibnamefont
  {Adler}},\ }\href@noop {} {\bibfield  {journal} {\bibinfo  {journal} {Bull.
  Am. Math. Soc.}\ }\textbf {\bibinfo {volume} {35}},\ \bibinfo {pages} {1}
  (\bibinfo {year} {1998})}\BibitemShut {NoStop}%
\bibitem [{\citenamefont {Mehta}(2004)}]{mehta}%
  \BibitemOpen
  \bibfield  {author} {\bibinfo {author} {\bibfnamefont {M.~L.}\ \bibnamefont
  {Mehta}},\ }\href@noop {} {\emph {\bibinfo {title} {Random Matrices}}}\
  (\bibinfo  {publisher} {{Elsevier, Amsterdam}},\ \bibinfo {year}
  {2004})\BibitemShut {NoStop}%
\bibitem [{\citenamefont {G\"orlich}\ and\ \citenamefont
  {Jarosz}(2004)}]{gorlich}%
  \BibitemOpen
  \bibfield  {author} {\bibinfo {author} {\bibfnamefont {A.}~\bibnamefont
  {G\"orlich}}\ and\ \bibinfo {author} {\bibfnamefont {A.}~\bibnamefont
  {Jarosz}},\ }\href@noop {} {} (\bibinfo {year} {2004}),\ \Eprint
  {http://arxiv.org/abs/arXiv:math-ph/0408019} {arXiv:arXiv:math-ph/0408019}
  \BibitemShut {NoStop}%
\bibitem [{\citenamefont {Sano}(2000)}]{sano00}%
  \BibitemOpen
  \bibfield  {author} {\bibinfo {author} {\bibfnamefont {M.~M.}\ \bibnamefont
  {Sano}},\ }\href@noop {} {\bibfield  {journal} {\bibinfo  {journal} {CHAOS}\
  }\textbf {\bibinfo {volume} {10}},\ \bibinfo {pages} {195} (\bibinfo {year}
  {2000})}\BibitemShut {NoStop}%
\bibitem [{\citenamefont {Garc\'ia-Mata}\ \emph {et~al.}()\citenamefont
  {Garc\'ia-Mata}, \citenamefont {Vallejos},\ and\ \citenamefont
  {Wisniacki}}]{pulpo_wip}%
  \BibitemOpen
  \bibfield  {author} {\bibinfo {author} {\bibfnamefont {I.}~\bibnamefont
  {Garc\'ia-Mata}}, \bibinfo {author} {\bibfnamefont {R.~O.}\ \bibnamefont
  {Vallejos}}, \ and\ \bibinfo {author} {\bibfnamefont {D.~A.}\ \bibnamefont
  {Wisniacki}},\ }\href@noop {} {}\bibinfo {note} {In preparation}\BibitemShut
  {NoStop}%
\end{thebibliography}
%
\end{document}